\documentstyle[12pt,epsfig]{article}

\catcode`\@=11
\def\@citex[#1]#2{%
\if@filesw \immediate \write \@auxout {\string \citation {#2}}\fi
\@tempcntb\m@ne \let\@h@ld\relax \def\@citea{}%
\@cite{%
  \@for \@citeb:=#2\do {%
    \@ifundefined {b@\@citeb}%
      {\@h@ld\@citea\@tempcntb\m@ne{\bf ?}%
      \@warning {Citation `\@citeb ' on page \thepage \space undefined}}%
      {\@tempcnta\@tempcntb \advance\@tempcnta\@ne%
      \@tempcntb\number\csname b@\@citeb \endcsname \relax%
      \ifnum\@tempcnta=\@tempcntb 
        \ifx\@h@ld\relax%
          \edef \@h@ld{\@citea\csname b@\@citeb\endcsname}%
        \else%
          \edef\@h@ld{\ifmmode{-}\else--\fi\csname b@\@citeb\endcsname}%
        \fi%
      \else
        \@h@ld\@citea\csname b@\@citeb \endcsname%
        \let\@h@ld\relax%
      \fi}%
    \def\@citea{,\penalty\@highpenalty\,}%
  }\@h@ld
}{#1}}

\def\@citeb#1#2{{[#1]\if@tempswa , #2\fi}}
\def\@citeu#1#2{{$^{#1}$\if@tempswa , #2\fi }}
\def\@citep#1#2{{#1\if@tempswa , #2\fi}}

\def\bcites{         
        \catcode`\@=11
        \let\@cite=\@citeb
        \catcode`\@=12
}

\def\upcites{         
        \catcode`\@=11
        \let\@cite=\@citeu
        \catcode`\@=12
}

\def\plaincites{      
        \catcode`\@=11
        \let\@cite=\@citep
        \catcode`\@=12
}

\newcount\hour
\newcount\minute
\newtoks\amorpm
\hour=\time\divide\hour by 60
\minute=\time{\multiply\hour by 60 \global\advance\minute by-\hour}
\edef\standardtime{{\ifnum\hour<12 \global\amorpm={am}%
        \else\global\amorpm={pm}\advance\hour by-12 \fi
        \ifnum\hour=0 \hour=12 \fi
        \number\hour:\ifnum\minute<10 0\fi\number\minute\the\amorpm}}
\edef\militarytime{\number\hour:\ifnum\minute<10 0\fi\number\minute}

\def\draftlabel#1{{\@bsphack\if@filesw {\let\thepage\relax
   \xdef\@gtempa{\write\@auxout{\string
      \newlabel{#1}{{\@currentlabel}{\thepage}}}}}\@gtempa
   \if@nobreak \ifvmode\nobreak\fi\fi\fi\@esphack}
        \gdef\@eqnlabel{#1}}
\def\@eqnlabel{}
\def\@vacuum{}
\def\marginnote#1{}
\def\draftmarginnote#1{\marginpar{\raggedright\scriptsize\tt#1}}
\def\draft{
        \pagestyle{plain}
        \overfullrule=2pt
        \oddsidemargin -.5truein
        \def\@oddhead{\sl \phantom{\today\quad\militarytime} \hfil
        \smash{\Large\sl DRAFT} \hfil \today\quad\militarytime}
        \let\@evenhead\@oddhead
        \let\label=\draftlabel
        \let\marginnote=\draftmarginnote
        \def\ps@empty{\let\@mkboth\@gobbletwo
        \def\@oddfoot{\hfil \smash{\Large\sl DRAFT} \hfil}
        \let\@evenfoot\@oddhead}
        \def\@eqnnum{(\theequation)\rlap{\kern\marginparsep\tt\@eqnlabel}%
        \global\let\@eqnlabel\@vacuum}  }

\def\blackfonts{
        \font\blackboard=msbm10 scaled\magstep1
        \font\blackboards=msbm8
        \font\blackboardss=msbm6
}

\def\nblack{            
        \def\ZZ{{Z \n{10} Z}}
        \def\NN{{N \n{14} N}}
        \def\CC{{C \n{11} C}}
        \def\RR{{R \n{11} R}}
        \def\QQ{{Q \n{12} Q}}
        \def\PP{{P \n{11} P}}
}

\def\prep{         
        \catcode`\@=11
        \input art10.sty
        \catcode`\@=12
        
        \let\small\null
        \def\blackfonts{
                \font\blackboard=msbm10
                \font\blackboards=msbm7
                \font\blackboardss=msbm5
        }
        \let\sl\it
        \twocolumn
        \sloppy
        \voffset=-2.54truecm
        \hoffset=-2.54truecm
        \flushbottom
        \parindent 1em
        \leftmargini 2em
        \leftmarginv .5em
        \leftmarginvi .5em
        \marginparwidth 48pt
        \marginparsep 10pt
        \setlength{\columnsep}{2truecm}
        \setlength{\textwidth}{25.4truecm}
        \setlength{\textheight}{17truecm}
        \baselineskip=16pt
        \oddsidemargin .18truein
        \evensidemargin .17truein
}

\def\eqalign#1{\null\,\vcenter{\openup\jot\m@th
  \ialign{\strut\hfil$\displaystyle{##}$&$\displaystyle{{}##}$\hfil
      \crcr#1\crcr}}\,}
\def\eqalignno#1{\displ@y \tabskip\centering
  \halign to\displaywidth{\hfil$\@lign\displaystyle{##}$\tabskip\z@skip
    &$\@lign\displaystyle{{}##}$\hfil\tabskip\centering
    &\llap{$\@lign##$}\tabskip\z@skip\crcr
    #1\crcr}}

\def\section{\@startsection {section}{1}{\z@}{3.ex plus 1ex minus
 .2ex}{2.ex plus .2ex}{\large\bf}}
\def\subsection{\@startsection{subsection}{2}{\z@}{2.75ex plus 1ex minus
 .2ex}{1.5ex plus .2ex}{\bf}}        

\def\appendix{{\newpage\section*{Appendix}}\let\appendix\section%
        {\setcounter{section}{0}
        \gdef\thesection{\Alph{section}}}\section}

\def\abstract{\if@twocolumn
\section*{Abstract}
\else 
\begin{center}
{\bf Abstract\vspace{-.5em}\vspace{0pt}}
\end{center}
\quotation
\fi}

\catcode`\@=12

\newcommand{\beq}{\begin{equation}}
\newcommand{\eeq}{\end{equation}}
\newcommand{\beqa}{\begin{eqnarray}}
\newcommand{\eeqa}{\end{eqnarray}}
\newcommand{\nn}{\nonumber}

\newcommand{\R}{{\bf R}}
\newcommand{\C}{{\bf C}}
\newcommand{\e}{{\rm e}}

\newcommand{\dd}{{\rm d}}

\newcommand{\ft}[2]{{\textstyle\frac{#1}{#2}}}
\newcommand{\eqn}[1]{(\ref{#1})}

\def\noj#1,#2,{{\bf #1} (19#2)\ }
\def\jou#1,#2,#3,{{\sl #1\/ }{\bf #2} (19#3)\ }
\def\ann#1,#2,{{\sl Ann.\ Physics\/ }{\bf #1} (19#2)\ }
\def\cmp#1,#2,{{\sl Comm.\ Math.\ Phys.\/ }{\bf #1} (19#2)\ }
\def\ma#1,#2,{{\sl Math.\ Ann.\/ }{\bf #1} (19#2)\ }
\def\ng#1,#2,{{\sl Nagoya.\ Math.\ J.\/ }{\bf #1} (19#2)\ }
\def\jd#1,#2,{{\sl J.\ Diff.\ Geom.\/ }{\bf #1} (19#2)\ }
\def\invm#1,#2,{{\sl Invent.\ Math.\/ }{\bf #1} (19#2)\ }
\def\cq#1,#2,{{\sl Class.\ Quantum Grav.\/ }{\bf #1} (19#2)\ }
\def\cqg#1,#2,{{\sl Class.\ Quantum Grav.\/ }{\bf #1} (19#2)\ }
\def\ijmp#1,#2,{{\sl Int.\ J.\ Mod.\ Phys.\/ }{\bf A#1} (19#2)\ }
\def\jmphy#1,#2,{{\sl J.\ Geom.\ Phys.\/ }{\bf #1} (19#2)\ }
\def\jams#1,#2,{{\sl J.\ Amer.\ Math.\ Soc.\/ }{\bf #1} (19#2)\ }
\def\grg#1,#2,{{\sl Gen.\ Rel.\ Grav.\/ }{\bf #1} (19#2)\ }
\def\mpl#1,#2,{{\sl Mod.\ Phys.\ Lett.\/ }{\bf A#1} (19#2)\ }
\def\nc#1,#2,{{\sl Nuovo Cim.\/ }{\bf #1} (19#2)\ }
\def\np#1,#2,{{\sl Nucl.\ Phys.\/ }{\bf B#1} (19#2)\ }
\def\pl#1,#2,{{\sl Phys.\ Lett.\/ }{\bf #1B} (19#2)\ }
\def\pla#1,#2,{{\sl Phys.\ Lett.\/ }{\bf #1A} (19#2)\ }
\def\pr#1,#2,{{\sl Phys.\ Rev.\/ }{\bf #1} (19#2)\ }
\def\prd#1,#2,{{\sl Phys.\ Rev.\/ }{\bf D#1} (19#2)\ }
\def\prl#1,#2,{{\sl Phys.\ Rev.\ Lett.\/ }{\bf #1} (19#2)\ }
\def\prp#1,#2,{{\sl Phys.\ Rept.\/ }{\bf #1C} (19#2)\ }
\def\ptp#1,#2,{{\sl Prog.\ Theor.\ Phys.\/ }{\bf #1} (19#2)\ }
\def\ptpsup#1,#2,{{\sl Prog.\ Theor.\ Phys.\/ Suppl.\/ }{\bf #1} (19#2)\ }
\def\rmp#1,#2,{{\sl Rev.\ Mod.\ Phys.\/ }{\bf #1} (19#2)\ }
\def\yadfiz#1,#2,#3[#4,#5]{{\sl Yad.\ Fiz.\/ }{\bf #1} (19#2) #3%
\ [{\sl Sov.\ J.\ Nucl.\ Phys.\/ }{\bf #4} (19#2) #5]}
\def\zh#1,#2,#3[#4,#5]{{\sl Zh.\ Exp.\ Theor.\ Fiz.\/ }{\bf #1} (19#2) #3%
\ [{\sl Sov.\ Phys.\ JETP\/ }{\bf #4} (19#2) #5]}

\def\beq{\begin{equation}}
\def\eeq{\end{equation}}
\def\beqar{\begin{eqnarray}}
\def\eeqar{\end{eqnarray}}

\newcommand{\be}{\begin{equation}}
\newcommand{\ee}{\end{equation}}
\newcommand{\bea}{\begin{eqnarray}}
\newcommand{\eea}{\end{eqnarray}}

\def\nfrac#1#2{{\displaystyle{\vphantom1\smash{\lower.5ex\hbox{\small$#1$}}%
        \over\vphantom1\smash{\raise.25ex\hbox{\small$#2$}}}}}

\def\n#1{\mskip-#1mu}

\def\to{\rightarrow}

\def\lae{\mathrel{\mathop{\smash{\lower .5 ex \hbox{$\stackrel<\sim$}}}}}
\def\lae{\mathrel{\mathop{\smash{\lower .5 ex \hbox{$\stackrel>\sim$}}}}}

\def\l:{\mathopen{:}\,}
\def\r:{\,\mathclose{:}}

\catcode`\@=11
\def\theequation{\arabic{equation}}
\catcode`\@=12

\nblack
\bcites

\nblack

\catcode`\@=11
\def\theequation{\thesection.\arabic{equation}}
\@addtoreset{equation}{section}
\@addtoreset{footnote}{section}
\@addtoreset{footnote}{subsection}
\catcode`\@=12

\newcommand{\beqn}{\begin{equation}}
\newcommand{\eeqn}{\end{equation}}
\newcommand{\beqnarray}{\begin{eqnarray}}
\newcommand{\eeqnarray}{\end{eqnarray}}
\newcommand {\bear} [1] {\begin {array} {#1}}
\newcommand {\ear} {\end {array}}

\newcommand {\beqarn} {\begin{eqnarray*}}
\newcommand {\eeqarn} {\end{eqnarray*}}

\newcommand{\cN}{{\cal N}}

\newcommand{\hSigma}{\widehat{\Sigma}}
\newcommand{\hgamma}{\widehat{\gamma}}
\newcommand{\he}{\widehat{e}}

\newcommand{\hQ}{\widehat{Q}}
\newcommand{\hR}{\widehat{R}}
\newcommand{\hphi}{\widehat{\phi}}
\newcommand{\hM}{\widehat{M}}
\newcommand{\sPhi}{\mbox{\small $\mit\Phi$}}
\newcommand{\Slin}{G}
\newcommand{\hSlin}{\widehat{G}}

\begin{document}

\begin{titlepage}

\begin{center}
\today
\hfill hep-th/0105075\\
\hfill HUTP-00/A020 \\ 
\hfill MIT-CTP-3100 \\ 
\hfill CU-TP-1004 \\

\vskip 1.5 cm
{\large \bf Mirror Symmetry in $2+1$ and $1+1$ Dimensions}
\vskip 1 cm 
{Mina Aganagic${}^1$, Kentaro Hori${}^1$, Andreas Karch${}^2$
and David Tong${}^3$}\\
\vskip 0.5cm
${}^1${\sl Jefferson Physical Laboratory, Harvard University, 
\\
Cambridge, MA 02138, U.S.A.\\}

\vskip 0.3cm
${}^2${\sl Center for Theoretical Physics, 
Massachusetts Institute of Technology, \\ Cambridge, MA 02139, U.S.A.\\}

\vskip 0.3cm
${}^3${\sl Department of Physics, Columbia University, \\ 
New York, NY 10027, U.S.A.\\}

\end{center}

\vskip 0.5 cm
\begin{abstract}
We study the Coulomb-Higgs duality of $\cN=2$ supersymmetric
Abelian Chern-Simons theories in $2+1$ dimensions, by compactifying
dual pairs on a circle of radius $R$
and comparing the resulting ${\cal N}=(2,2)$ theories in $1+1$ dimensions.
Below the compactification scale, the theory on the Higgs branch
reduces to the non-linear sigma model on a toric manifold.
In the dual theory on the Coulomb branch, the Kaluza-Klein modes
generate an infinite tower of contributions to the superpotential. 
After resummation, in the limit $R\to 0$
the superpotential becomes
that of the Landau-Ginzburg model which is the two-dimensional mirror
of the toric sigma model.
We further examine the conjecture of all-scale three-dimensional
mirror symmetry and observe that
it is consistent with mirror symmetry in $1+1$ dimensions.

\end{abstract}

\end{titlepage}

\section{Introduction}

Many of the dualities of interacting quantum field theories
exchange Noether charges and topological charges and
can be considered as generalizations of the Abelian duality
of free field theories. However, while Abelian dualities in various dimensions
are related by dimensional reduction, the same is 
not true of interacting theories. 
One reason for this is that non-trivial dualities are usually 
an infra-red equivalence while compactification probes
short-distance scales.
In this paper, we present one example where non-trivial dualities
are related by compactification. The dualities in question are the 
mirror symmetry of $(2,2)$ supersymmetric field theories
in $1+1$ dimensions \cite{HV}
and  the Coulomb-Higgs duality of supersymmetric gauge theories in 
$2+1$ dimensions, which is also known as mirror symmetry \cite{IS}.

Mirror symmetry of $(2,2)$ theories in $1+1$ dimensions
can be considered as a generalization of scalar-scalar
duality (T-duality).
It exchanges the vector and axial $U(1)$ R-symmetries, together 
with the symplectic geometric and the complex analytic aspects of 
the theories. A proof this duality for a class of theories
was recently offered in \cite{HV}: one starts 
with a gauged linear sigma-model and performs T-duality on 
each charged chiral multiplet.
 The new, dual, variables that one obtains 
are naturally adapted for describing instantons which, in two 
dimensional gauge theories, are Nielsen-Olesen vortices.
After accounting for such effects, the resulting mirror theory
is a Landau-Ginzberg (LG) model with a Toda-type superpotential.

On the other hand, mirror symmetry of gauge theories in $2+1$ dimensions 
was originally proposed for theories with ${\cal N}=4$ supersymmetry 
\cite{IS}, and can be thought of as a generalization of scalar-vector 
duality. It exchanges the two $SU(2)$ R-symmetry groups
(hence the name ``mirror symmetry''),
and correspondingly, the Coulomb and the Higgs branches.
This duality is not as well understood as the two-dimensional mirror
symmetry in the sense that no derivation has been given 
except via embeddings into string theory \cite{stringy}, 
which involves the subtle issue of decoupling extra degrees of freedom.
Nevertheless there exists considerable field theoretic
evidence. Moreover, there exists a tantalising 
similarity with the transformation used in \cite{HV}:
mirror symmetry in three dimensions exchanges
the bound states of electrons with Nielson-Olesen vortices \cite{ahiss}. 
One may therefore wonder whether there exists a deeper connection between 
the two dualities.

In order to make 
more quantitative contact with the two-dimensional theories,
we must firstly flow to 
three-dimensional mirror pairs with ${\cal N}=2$ supersymmetry. 
A prescription for this was given in \cite{Dave}
by weakly gauging a diagonal combination of the R-symmetry currents.
This induces mass splittings for the hypermultiplets which, 
in turn, leads to the dynamical generation of Chern-Simons couplings.
For Abelian theories, the resulting Chern-Simons
mirror pairs were previously analyzed in \cite{nd}
where it was shown that the Coulomb and Higgs branches
do indeed coincide as toric varieties.

In this paper, we compactify the mirror pairs of \cite{nd,Dave} 
on a circle of radius $R$ and study the two-dimensional ``continuum'' limit 
$R\to 0$. 
We tune the parameters of one theory (on the Higgs branch) to ensure that 
it descends to a two-dimensional non-linear sigma model on a toric manifold.
The corresponding dual theory (on the Coulomb branch) is 
then analyzed in the same limit, where it appears as a $1+1$ dimensional gauge
theory with infinitely many Kaluza-Klein (KK) matter fields.
Each of the charged KK modes generates a superpotential term
on the Coulomb branch, computed exactly at the one-loop level,
and we sum this infinite tower of terms. 
In the limit $R\to 0$, we obtain the superpotential for the LG mirror
\cite{HV} of the toric sigma model.

Thus, we show that three-dimensional mirror symmetry descends, upon
compactification, to two-dimensional mirror symmetry.
This may come as something of a surprise since
the former is originally regarded as an infra-red duality.
There is, however, no surprise: the superpotential
is a holomorphic quantity and 
is independent of the ratio of the compactification scale and the
scale set by the three-dimensional gauge coupling. Equipped with 
an infra-red duality, one would therefore expect to be able to re-derive 
agreement between such holomorphic quantitites, while non-holomorphic objects, 
such as the Kahler potential, would be beyond our reach. However, there is a 
proposal \cite{ks} that, after a suitable modification of the theories, 
three-dimensional mirror symmetry holds beyond the infra-red limit.
Given this proposal, we also attempt a comparison of the Kahler potential, 
and indeed find agreement with two-dimensional mirror symmetry
\cite{HV,HK}. This observation can be regarded as support for the
3d duality at all length scales.

The rest of the paper is organized as follows.
In Section~\ref{sec:pre},
we warm up by recalling some basic facts on compactification
from $2+1$ to $1+1$ dimensions.
Section~\ref{sec:com} is the main part of this paper.
We study the compactification of the Model A/Model B
mirror pairs of \cite{nd,Dave}.
We tune the parameters so that Model A reduces to
the non-linear sigma model on a toric manifold. For Model B, 
we compute the exact superpotential after compactification 
and show that in the limit $R\to 0$ it becomes the LG 
superpotential of the 2d mirror.
In Section~\ref{sec:int}, we further study the Kahler potential
of the all-scale mirror pairs proposed in \cite{ks}, which requires 
a modification of Model A. Upon compactification we obtain 
two-dimensional mirror symmetry between the sigma model on a sqashed
toric manifold and the LG model with a finite Kahler potential, in 
agreement with two-dimensional mirror symmetry.
In Section~\ref{sec:vor}, we discuss the BPS states of the three-dimensional
mirror pairs and, in particular, the relationship to the 
compactification analysis. In Section~\ref{conclude} we conclude and 
outline some directions for future research. We also include 
two appendices. In Appendix~\ref{app:A}, we specify the regularization 
scheme we are using in this paper and resolve a subtle issue in the
compactification analysis of Section~\ref{sec:com}.
In Appendix~\ref{app:B}, we review the Coulomb branch analysis of
\cite{nd} and strengthen the statement by proving a  non-perturbative
non-renormalization theorem of the Kahler class.

\section{Compactification Preliminaries}
\label{sec:pre}

In this section we discuss general aspects of
${\cal N}=2$ supersymmetric
gauge theories in $2+1$ dimensions compactified on a circle $S^1$
of radius $R$. (Such compactifications
were considered, for example, in \cite{ds,ss,ak}.)
We denote the space-time coordinates by $x^0,x^1,x^2$ where
$x^0,x^1$ are the time and space coordinates of the infinite
$1+1$ dimensions and $x^2$ is the coordinate of the circle $S^1$,
with period $2\pi R$.

The relationship between the two and three-dimensional coupling
constants is,
\beq
{2\pi R\over e_{\rm 3d}^2}={1\over e_{\rm 2d}^2}.
\eeq
In general the physics of the system depends on the dimensionless
combination
\beq
\gamma=2\pi R\,e^2_{\rm 3d}.
\eeq
For $\gamma \gg 1$, the system flows first 
to infra-red three-dimensional physics, before flowing to 
two dimensions. In contrast, for $\gamma\ll 1$, the system is 
essentially two-dimensional by the time
strong three-dimensional gauge interactions play a role. 

From the point of view of two-dimensions, a three-dimensional
field consists of a tower of infinitely many Kaluza-Klein
fields, each of which is a mode of the Fourier expansion
in the $x^2$ direction.
The Kaluza-Klein fields of a
three-dimensional chiral multiplet of real mass $m$
are all two-dimensional chiral multiplets, with 
twisted masses given by $(m+i n/R)$, for integer $n$. 
A three-dimensional Abelian vector multiplet contains a single real
scalar $\phi$, a Dirac fermion, and a $U(1)$ gauge field $v_\mu$,
together with an auxillary scalar $D$. 
Its Kaluza-Klein modes are
a two-dimensional vector multiplet $V$, 
together with an infinite tower of massive Kaluza-Klein modes.
The two-dimensional vector multiplet contains a complex scalar
field which decomposes into $\sigma=\sigma_1+i\sigma_2$ 
with
\beqa
\sigma_1&=&\frac{1}{2\pi R}\int_{S_1}\phi \nn\\
\sigma_2&=&\frac{1}{2\pi R}\int_{S_1}v\equiv\sigma_2+\frac{1}{R}\,.
\label{perio}
\eeqa
where the periodicity of the Wilson line $\sigma_2$ arises from large 
gauge transformations.
Other fields in $V$ descend by obvious dimensional 
reduction from three-dimensions.
They can be combined into a gauge invariant 
twisted chiral superfield $\Sigma= \overline{D}_+D_-V$
whose lowest component is $\sigma$.
The massive vector Kaluza-Klein modes are also
twisted chiral multiplets whose scalar components are
the non-constant Fourier modes of $\phi$ and the Wilson line.
These multiplets have twisted masses $in/R$ for integer $n$.

\subsection{Fayet-Iliopoulos and Chern-Simons Couplings}

The three-dimensional FI parameter $\zeta$ and the two-dimensional
FI parameter $r$ are related by
\beq
r=2\pi R \,\zeta
\label{myfi}\eeq
In the two-dimensional theory, the FI coupling is given by a twisted 
superpotential
\beq
\widetilde{W}_{FI}=-r\Sigma=-2\pi R\zeta\Sigma
\label{whassupnow}
\eeq
Let us see how the CS coupling is described in two-dimensions.
Its bosonic part is given by
\beq
CS=\frac{k}{4\pi}\int\limits_{\R^2\times S^1}
\left(v\wedge \dd v +2\phi D\,\dd^3x\right) \ .
\eeq
For constant modes the intergal becomes two-dimensional
with the integrand given by 
$Rk\ {\rm Re}\{\sigma (D-iv_{01})\}$.
The supersymmetric completion amounts to
the twisted superpotential
\beq
\widetilde{W}_{CS}=-\pi Rk \Sigma^2.
\label{WHASSSUP}
\eeq
Alternatively, we may derive this
entirely within superspace: the supersymmetric 
Chern-Simons interaction is given in terms of the linear superfield
$\Slin=\epsilon^{\alpha\beta}\overline{D}_{\alpha}D_{\beta}V$.
The dimensional reduction of this superfield may be expressed
in terms of a twisted superfield,
$\Slin=\Sigma+\overline{\Sigma}$. The Chern-Simons 
interaction now reads
\beq
CS_{\rm SUSY}\,=\,
-{k\over 4\pi}\int_{\R^2\times S^1}\dd^3x\dd^4\theta\, \Slin V
\,=\,
-{2\pi Rk\over 4\pi}
{\rm Re}\int_{\R^2}\dd^2x\dd^2\widetilde{\theta}
\,\,\Sigma^2
\eeq
in agreement with the twisted superpotential \eqn{WHASSSUP}.

\subsection{Abelian Duality in $2+1$ and $1+1$ dimensions}
\label{subsec:ab}

In $2+1$ dimensions, a $U(1)$ gauge field is dual to a
compact scalar field.
Let us see how this duality looks like when compactified on a circle.
We consider here the simplest free bosonic theory.
In the free theory, the massive Kaluza-Klein modes
are not coupled to the zero mode and simply go away
in the limit $R\to 0$.

The three-dimensional duality is between a scalar field
$\varphi$ of period $2\pi$ with the action
\beq
S_{3d}={1\over 2\pi}\int \dd^3x {\lambda^2\over 2}\left(
(\partial_0\varphi)^2-(\partial_1\varphi)^2-(\partial_2\varphi)^2\right),
\eeq
and a $U(1)$ gauge field $v_{\mu}$ with the action
\beq
\widehat{S}_{3d}={1\over 2\pi}
\int \dd^3x {1\over 2\lambda^2}
\left((v_{01})^2+(v_{02})^2-(v_{12})^2\right).
\eeq
The gauge coupling of the dual theory is
$\lambda$, and thus the $\gamma$ parameter of the compactification
is given by
\beq
\hgamma=2\pi R\lambda^2.
\eeq
The compactification of the scalar field theory reduces in the
$R\to 0$ limit to the $1+1$ dimensional theory
with the action
\beq
S_{2d}={1\over 2\pi}\int \dd^2x {\hgamma\over 2}\left(
(\partial_0\varphi)^2-(\partial_1\varphi)^2\right).
\label{S2}
\eeq
This is the action for a sigma model on a circle of radius
$\sqrt{\hgamma}$.
For the dual gauge theory, we note that the Wilson line $\sigma_2$
(\ref{perio}) has a finite periodicity after the rescaling
$\vartheta:=2\pi R \sigma_2\equiv \vartheta+2\pi$.
In terms of the rescaled variable, the compactified theory
reduces in the limit $R\to 0$ as
\beq
\widehat{S}_{2d}={1\over 2\pi}\int \dd^2x
{1\over 2\hgamma}\left((\partial_0\vartheta)^2-(\partial_1\vartheta)^2
+(2\pi Rv_{01})^2\right).
\label{tS2}
\eeq
In $1+1$ dimensions, a Maxwell field has no propagating modes,
and also, it has no finite energy topological excitation
in the limit $\sqrt{\hgamma}/R\to \infty$.
Thus, the system (\ref{tS2}) describes the sigma model on a circle of
radius $1/\sqrt{\hgamma}$, which is T-dual to (\ref{S2}).

We have seen that the Abelian duality in $2+1$ dimensions descends
to the Abelian duality in $1+1$ dimensions.
In the language of the gauge theory,
T-duality is between
the dual photon $\varphi$ and the Wilson line $\vartheta$.
In the following, we shall see an analogous phenomenon
in interacting, supersymmetric theories.
Unlike in the free theory however, the massive Kaluza-Klein
modes play an important role.

\section{Compactification of 3D Mirrors}
\label{sec:com}

In this section, which is the main part of this paper,
we consider compactification of a mirror pair of interacting
${\cal N}=2$ gauge theories in $2+1$ dimensions.
We will tune the parameters (as a function of the compactification scale
$1/2\pi R$) so that one theory reduces to the $1+1$ dimensional
non-linear sigma model
on the Higgs branch.
Then, we will see that the other theory on the Coulomb branch
reduces to the Landau-Ginzburg
model in $1+1$ dimensions,
which is the 2d mirror \cite{HV} of the sigma model on the Higgs branch.

\subsection{Aspects of the 3D Models}

We consider mirror pairs of $\cN=2$ Abelian Chern-Simons gauge theories
found  in \cite{nd}.

\noindent
{\bf Model A:}\ $U(1)^k$ gauge group with $N$ chiral multiplets, $\Phi_i$, 
$(i=1,\ldots,N)$ of charge $Q_i^a$  under the $a^{\rm th}$ $U(1)$ 
factor $(a=1,\ldots,k)$. The Chern-Simons couplings are given by 
$k^{ab}={1\over 2}\sum_{i=1}^NQ^a_{i}Q^b_{i}$. 
The theory is parametrized by the gauge coupling constants $e_a$,
the FI parameters $\zeta^a$ and the real masses $m_i$ for $\Phi_i$.

\newcommand{\hPhi}{\widehat{\Phi}}
\newcommand{\hz}{\widehat{\zeta}}
\newcommand{\hm}{\widehat{m}}
\newcommand{\blambda}{\overline{\lambda}}
\newcommand{\btheta}{\overline{\theta}}
\newcommand{\sTheta}{\mbox{\small $\sl\Theta$}}
\newcommand{\sDelta}{{\mit\Delta}}

\noindent
{\bf Model B:}\ $U(1)^{N-k}$ gauge group with $N$ chiral multiplets
$\hPhi_i$, $(i=1,\ldots,N)$ of charge $\hQ_i^p$ under the 
$p^{\rm th}$ $U(1)$ factor $(p=1,\ldots,N-k)$.
The Chern-Simons couplings are given by
$\widehat{k}^{pq}=-{1\over 2}\sum_{i=1}^N\hQ_i^p\hQ_{i}^q$. 
The theory is parametrized by the gauge coupling constants
$\widehat{e}_p$,
the FI parameters $\hz^{p}$ and the real masses $\hm_i$ 
for $\hPhi_i$.

\noindent
The two sets of charges obey
\beq
\sum_{i=1}^NQ^a_{i}\hQ_i^p=0,~~\forall\ a~~{\rm and}~~\forall\ p.
\eeq
The mass and the FI parameters are related by the mirror map
\beq
\begin{array}{l}
{\displaystyle
\zeta^a-{1\over 2}\sum_{i=1}^N Q_i^am_i=\sum_{i=1}^NQ_i^a\hm_i},
\\[0.2cm]
{\displaystyle
-\sum_{i=1}^N\hQ_i^pm_i=\hz^p+{1\over 2}\sum_{i=1}^N\hQ_i^p\hm_i.}
\end{array}
\label{mm}
\eeq
Of the $(N+k)$ mass and FI parameters describing Model A, only $N$ are
physical. This is because a shift of the scalar fields in the vector
multiplets may be compensated by a shift of the parameters
\beq
m_i\to m_i+\sum_{b=1}^kQ_i^bc_b,~~~
\zeta^a\to \zeta^a+\sum_{b=1}^kk^{ab}c_b.
\label{shift}\eeq
A similar remark applies to Model B where, once again, only $N$ out of the 
$(2N-k)$ mass and FI parameters are physical. The mirror map (\ref{mm})
is the relation between these physical parameters.

The theories are not finite and
the FI parameters have to be renormalized.
Therefore, the mirror map (\ref{mm})
between FI and mass parameters depends
on the regularization scheme.
In fact, the mirror pairs \cite{dhooy} of finite ${\cal N}=4$ theories
can serve as cut-off theories,
and a specific regularization scheme is chosen
to find the mirror map (\ref{mm}).
The details are recorded in Appendix \ref{app:A}.

The Higgs branch of Model A
corresponds to the Coulomb branch of Model B.
While the Higgs branch is determined at the classical 
level by a symplectic quotient,
the structure of the Coulomb branch arises due to  
certain quantum effects particular to $2+1$ dimensions.
Specifically, there exist massless photons if,
after integrating out the massive chiral multiplets, 
the effective Chern-Simons coefficients vanish.
This defines the base of the Coulomb branch.
The dual photons then provide the torus fibration.
At the full quantum level,
the Higgs branch of Model A
and the Coulomb branch of Model B
are exactly the same toric variety \cite{nd} 
with the same K\"ahler class, as shown in Appendix \ref{app:B}.

Unlike in theories with ${\cal N}=4$ supersymmetry, integrating out
the chiral multiplets leads to a finite renormalization 
of the FI parameter.
Thus, while the full Higgs branch exists only for vanishing masses, the 
Coulomb branch exists for vanishing ${\it effective}$ FI parameter. In 
the region in which the Chern-Simons coupling vanishes, the effective FI 
parameter is the following physical combination
\beq
\begin{array}{l}
\zeta^a_{\rm eff}=\zeta^a-\frac{1}{2}\sum_{i=1}^NQ^a_im_i\\[0.1cm]
\hz^p_{\rm eff}=\hz^p+\frac{1}{2}\sum_{i=1}^N\hQ^p_i\hm_i
\end{array}
\label{zetaeff}
\eeq
where the $\pm$ signs may be traced back to the $\pm$
signs of the  Chern-Simons couplings. Notice that it is these effective 
FI parameters which appear in the mirror map (\ref{mm}).

In the past, little attention has been paid to the normalization of the 
mirror map. However, this will prove to be crucial for our story. 
We use conventions in which the auxiliary D-fields appear in the action
as ${1\over 2\pi}\int\dd^3x({1\over 2e^2}D^2-\zeta D)$, as in \cite{HV}.
Then, (\ref{mm}) is the correct one.
This can be most easily seen by comparing the masses of
the BPS vortices at special points of the Higgs branch 
with the masses of BPS electrons at special 
points of the Coulomb branch and recalling that these
states are exchanged under
mirror symmetry \cite{ahiss}.
The check, with the sign,
can also be made using the manipulation in \cite{ks}.

\subsection{Compactification of Model A on the Higgs branch}

We compactify Model A on the circle of radius $R$ so that at energies
below $1/R$ we obtain the supersymmetric gauge theory
in $1+1$ dimensions with the same gauge group and field content.
We will focus on the Higgs branch of the model.

The FI parameters in two-dimensional gauge theories are renormalized
as
\beq
r^a(\mu')=r^a(\mu)+\sum_{i=1}^NQ_{i}^{a}\log(\mu'/\mu).
\label{rrun}
\eeq
This applies, in the present case,
only at energies below the compactification scale
\beq
\Lambda_{\rm UV}=1/2\pi R.
\eeq
Thus, in order to have the renormalized two-dimensional theory,
the three-dimensional FI parameters must depend on the radius
$R$ as
\beq
\zeta^a={1\over 2\pi R}r^a(\Lambda_{\rm UV})
={1\over 2\pi R}\left(\sum_{i=1}^NQ_i^a\log(1/2\pi R\Lambda)+r^a\right),
\label{ascale}
\eeq
where the scale $\Lambda$
and parameters $r^a$ are taken fixed in the continuum limit
$1/R\to \infty$ of the two-dimensional theory.
If $\sum_{i=1}^NQ_i^a\ne 0$ for some $a$,
$\Lambda$ is a physical RG invariant scale parameter
of the theory which replaces one combination of $r^a$
(for $k=1$ it is standard to take $r=0$).
If $\sum_{i=1}^NQ_i^a=0$ for all $a$,
all $r^a$'s are physical parameters of the theory.
\footnote{We could dispense with these remarks
if we wrote (\ref{ascale})
as $\zeta^a
={1\over 2\pi R}(\sum_{i=1}^NQ_i^a\log(1/2\pi R\mu)+r^a(\mu))$
as usual. Instead, we express it in terms of the
RG invariant scale $\Lambda$
(and parameters $r^a$) since the running 2d coupling may be confusing
in what follows.}

We now set the FI parameters to be sufficiently large and all the real masses
to zero. This ensures that, at low enough energies,
the theory is the non-linear sigma model on the Higgs branch in
$1+1$ dimensions.
The Higgs branch is a toric manifold
obtained as the symplectic quotient
of $U(1)^k$ acting on $\C^N$ with charge $Q_i^a$. Namely,
the vacuum manifold
\beq
\sum_{i=1}^NQ_{i}^a|\sPhi_i|^2=\zeta^a
\eeq
modded out by the $U(1)^k$ gauge group action.
In the above expression, $\sPhi_i$ is the scalar component
of the three-dimensional chiral multiplet $\Phi_i$. In particular,
these scalars have classical vacuum expectation
values of order $\sqrt{\zeta}$ or
\beq
|\sPhi|^2\sim \zeta\sim {1\over R}
\label{region}
\eeq
In the above discussion we have implicitly assumed $\gamma=e^2_{3d}R$
to be small. 
However, this is not mandatory. We may take $\gamma\gg 1$,
in which case the theory 
reduces to the non-linear sigma model above the compactification scale 
$\Lambda_{\rm UV}=1/2\pi R$. As we flow to energies below $\Lambda_{\rm UV}$
the metric starts to run. The FI parameters play the role of the Kahler
class parameters which are known to be renormalized
only at the one loop level \cite{acg} (see also
\cite{hv2} for a simple derivation). This leads once again to the same $R$ 
dependence as \eqn{ascale}, for $\zeta^a$.

\subsection{Compactification of Model B on the Coulomb Branch}
\label{subsec:cptB}

We turn now to the compactification of Model B on the Coulomb branch.
We consider the limit $\hgamma:=\he^2_{3d}R\ll 1$ such that the theory flows 
to $1+1$ dimensional gauge theory before $2+1$ dimensional gauge 
interactions become important.
Thus, we will consider this theory as a gauge theory in $1+1$ dimensions.
We will see that the Kaluza-Klien modes,
considered as infinitely many charged and neutral matter fields, play an
important role.

Firstly we determine the parameters of Model B using the mirror
map \eqn{mm}. The 
vanishing of the real masses of Model A requires us to set the effective FI 
parameters to zero, ensuring that Model B has a Coulomb branch,
\be
\hz^p=-{1\over 2}\sum_{i=1}^N\hQ_i^p\hm_i
\label{hz}
\ee
From \eqn{ascale}, we require
the real masses of Model B to depend on the radius $R$
as
\beq
\widehat{m}_i={1\over 2\pi R}\Bigl(\log(1/2\pi R\Lambda)+r_i\Bigr)
\label{mi}
\eeq
where $r_i$ solves the equation $\sum_{i=1}^NQ_i^ar_i=r^a$.

We next discuss the appropriate variables to describe the Coulomb branch of 
Model B. Recall that the scalar components, 
$\hat{\phi}$ of the Model B vector multiplets are related to 
the scalar components $\sPhi$ of the Model A chiral multiplets as
\cite{IS,DHOO}
\beq
\widehat{\phi}\sim |\sPhi|^2\,.
\eeq
Comparing with (\ref{region}) we should rescale the Coulomb branch 
variables by $1/R$. Together with the supersymmetric partner, we rescale the 
superfield strength $\hSigma_p$ as
\beq
\hSigma_p=\Theta_p/2\pi R.
\label{resc}
\eeq
This is the same scaling we performed in the free theory of Section 
\ref{subsec:ab}. As in that case, the periodicity of
the imaginary part of $\hSigma_p$ given in 
(\ref{perio}) reveals
\beq
{\rm Im}\,\Theta_p\equiv {\rm Im}\,\Theta_p+2\pi.
\eeq
With this choice of the fields,
the tree level twisted superpotential
coming from the CS and the FI terms is expressed as
\beqa
\widetilde{W}_{CS}+\widetilde{W}_{FI}
&=&-\pi R\sum_{p,q}\widehat{k}^{pq}\hSigma_p\hSigma_q
-2\pi R\sum_p\hz^p\hSigma_p
\nn\\
&=&
{1\over 8\pi R}\sum_{i,p,q}\hQ_i^p\hQ_i^q\Theta_p\Theta_q
+{1\over 2}\sum_{i,p}\hQ_i^p\hm_i\Theta_p,
\label{WCSh}
\eeqa
where we have used
$\widehat{k}^{pq}=-(1/2)\sum_{i=1}^N\hQ_i^p\hQ_i^q$ and (\ref{hz}).

We consider the theory at energy $E$ much smaller than the compactification
scale $1/R$. Naively one may expect the effect of all the massive KK modes 
to disappear in the extreme two-dimensional limit $ER\to 0$. However, as we 
now see, the rescaling of fields (\ref{resc}) and masses \eqn{mi} 
ensures that this is not the case. 

Let us first integrate out all the KK modes of the chiral multiplets.
This can be done exactly by one-loop integral
because the action is quadratic in these variables.
The effect includes the generation of
the standard $\Sigma\log(\Sigma)$ type
twisted superpotential.
Summing up all such modes we obtain
\beq
\sDelta\widetilde{W}=-\sum_{i=1}^N\sum_{n=-\infty}^{\infty}
(\hSigma_i+in/R)
\Bigl(\log(\hSigma_i+in/R)-1\Bigr)
\label{summ}
\eeq
where
\beq
\hSigma_i=\sum_{p=1}^{N-k}\hQ_i^p\hSigma_p+\widehat{m}_i.
\eeq
We still have to integrate out the non-zero modes of
the three-dimensional vector multiplets and the high frequency modes
of $\hSigma_p$.
We claim that this does not induce any twisted superpotential.
Note that the effective perturbation
expansion parameters are given by
$e_{2d}^2/\Sigma^2\sim \hgamma/\Theta^2$, which is small
as long as $\hgamma\ll 1$ (and $\Theta$ finite).
Then any quantum correction depends on $\hgamma$
but it cannot enter into the twisted
superpotential since the gauge coupling $e_{2d}$
is not a twisted chiral parameter.
(There could also be tree-level corrections from elimination of
heavy fields by the equations of motion. 
However, conservation of momentum in the compactified dimension ensures 
that the equations of motion are solved by setting
these fields to zero.)
Thus, (\ref{summ}) is the exact quantum correction to the
twisted superpotential.

To perform the infinite sum, we firstly differentiate,
\beqa
{\partial\sDelta\widetilde{W}\over\partial\hSigma_p}
&=&-\sum_{i,n}\hQ_i^p\log(\hSigma_i+in/R)
\nonumber\\
&=&-\sum_{i=1}^N\hQ_i^p\log\left[2\pi R\hSigma_i
\prod_{n=1}^{\infty}\left(1+{R^2\hSigma_i^2\over n^2}\right)\right]
-\sum_{i=1}^N\hQ_i^p\log\left[{1\over 2\pi R}\prod_{n=1}^{\infty}
\left({n^2\over R^2}\right)\right].~~~
\label{infpro}
\eeqa
The infinite product in the first term can be performed, resulting in the 
argument of the logarithm equal to $2\sinh (\pi R\hSigma_i)$. 
The second term appears divergent, but it should be considered
to vanish in our specific regularization scheme
that leads to (\ref{mm}). This is explained in detail
in Appendix \ref{app:A}. 
(If we took another regularization, 
the mirror map (\ref{mm}) would contain a divergent term
and this would cancel the second term of (\ref{infpro}) in the end.
See Appendix \ref{app:A}.)
Thus, we obtain
\beq
{\partial\sDelta\widetilde{W}\over\partial\hSigma_p}
=-\sum_{i=1}^N\hQ_i^p\log(\e^{\pi R\hSigma_i}-\e^{-\pi R\hSigma_i}).
\label{dWS}
\eeq
In terms of the rescaled variable $\Theta_p$ defined by (\ref{resc})
we have
\beq
{\partial\sDelta\widetilde{W}\over\partial\Theta_p}
=-{1\over 2\pi R}
\sum_{i=1}^N\hQ_i^p\log(\e^{\Theta_i/2}
-\e^{-\Theta_i/2}),
\label{goodinnit}\eeq
where
\beq
\Theta_i=2\pi R\hSigma_i
=\sum_{q=1}^{N-k}\hQ_i^q\Theta_q+2\pi R\widehat{m}_i.
\eeq
Now, we recall from (\ref{mi}) 
that $2\pi R\widehat{m}_i$ is logarithmically divergent
as $R\to 0$.
Thus, within the argument of the logarithm, $\e^{\Theta_i/2}$ dominates over 
$\e^{-\Theta_i/2}$ and we have the expansion
\beqa
{\partial\sDelta\widetilde{W}\over\partial\Theta_p}
&=&-{1\over 2\pi R}
\sum_{i=1}^N\hQ_i^p\left\{{\Theta_i\over 2}
+\log(1-\e^{-\Theta_i})\right\}
\nonumber\\
&=&-{1\over 2\pi R}\sum_{i=1}^N\hQ_i^p
\left\{{\Theta_i\over 2}-\e^{-\Theta_i}+O((\e^{-\Theta_i})^2)\right\}.
\eeqa
Since the equation \eqn{mi} shows
\beq
\e^{-\Theta_i}
=2\pi R\Lambda\,\e^{-\sum_{q=1}^{N-k}\hQ_i^q\Theta_q-r_i},
\eeq
the higher order terms 
${1\over 2\pi R}O((\e^{-\Theta_i})^2)$
vanish in the limit $R\to 0$.
Thus, in this limit we obtain
\beq
{\partial\sDelta\widetilde{W}\over\partial\Theta_p}
=\sum_{i=1}^N\hQ_i^p
\left\{-{1\over 4\pi R}\sum_{q=1}^{N-k}\hQ_i^q\Theta_q
-{1\over 2}\hm_i
+\Lambda\,\e^{-\sum_{q=1}^{N-k}\hQ_i^q\Theta_q-r_i}
\right\}.
\eeq
This can be integrated as
\beq
\sDelta\widetilde{W}=-{1\over 8\pi R}\sum_{i,p,q}
\hQ_i^p\hQ_i^q\Theta_p\Theta_q
-{1\over 2}\sum_{i,p}\hQ_i^p\hm_i\Theta_p
-\Lambda \sum_{i=1}^N\e^{-\sum_{q=1}^{N-k}\hQ_i^q\Theta_q-r_i}.
\label{sDelW}
\eeq
The terms quadratic and linear in $\Theta_p$ are cancelled by the 
tree level terms (\ref{WCSh}). The total twisted superpotential is therefore 
given by
\beqa
\widetilde{W}_{\rm total}&=&\widetilde{W}_{CS}
+\widetilde{W}_{FI}+\sDelta\widetilde{W}
\nn\\
&=&
-\Lambda \sum_{i=1}^N\e^{-\sum_{p=1}^{N-k}\hQ_i^p\Theta_p-r_i}.
\label{Wtilde}
\eeqa
This is precisely the LG superpotential obtained in 
\cite{HV} for the mirror of the non-linear sigma model with toric target space.

\subsection*{\it Inclusion of Twisted Mass}

One may consider weakly gauging a flavour 
symmetry of Model A to introduce non-zero real masses $m_i$. In the 
infra-red, this deforms the sigma-model 
on the Higgs branch by introducing a potential proportional to the length 
squared of a holomorphic Killing vector associated to the flavour 
symmetry. It is a simple matter to track this deformation in Model B: 
It shifts $\hm_i$ and $\hz^p$ by a finite amount.
Change in $\e^{-2\pi R\hm_i}$ due to a finite shift of $\hm_i$
does not matter in the limit $R\to 0$.
The expression (\ref{WCSh}) of $\widetilde{W}_{CS}+\widetilde{W}_{FI}$
in terms of $\Theta_p$ and new $\hm_i$ is shifted by
$\sum_{i,p}\hQ_i^pm_i\Theta_p$.
Thus, it fails to cancel the $\Theta$-linear term from
$\sDelta\widetilde{W}$.
The total twisted superpotential is therefore given by
\beq
\widetilde{W}
=-\Lambda\sum_{i=1}^N
\e^{-\sum_{p=1}^{N-k} \hQ_i^p\Theta_p-r_i}
+\sum_{i,p}\hQ_i^pm_i\Theta_p
\eeq
once again in agreement with \cite{HV}.

\section{Interpretation}
\label{sec:int}

In the previous section we examined the fate of
three-dimensional field theories upon compactification on a 
circle. We found that the three-dimensional mirror pairs
lead to two-dimensional mirror pairs.
This may appear to be a consistent picture, but
one should note the following.
Three-dimensional mirror symmetry was originally conjectured
to be an infra-red duality, applying only 
in the limit $e_{3d},\widehat{e}_{3d}\rightarrow\infty$.
Equivalence of the two compactified theories
would hold, based on this conjecture,
only when the compactification scale $1/R$ is much smaller
than the scales $e_{3d}^2$ and $\widehat{e}_{3d}^2$,
namely only for $\gamma\gg 1,\hgamma\gg 1$.
However, we have considered the regime
$\hgamma\ll 1$ in Model B compactification,
but still we found an agreement with Model A compactification.
In this section, we discuss the meaning of this observation.

\subsection{No phase transition between $\hgamma\gg 1$ and $\hgamma\ll 1$
?}

Firstly, let us shift perspective slightly and repeat the calculations 
presented above, with a somewhat different philosophy. To this 
end, we consider only Model B on the Coulomb branch and examine two different 
descriptions of this theory: one in terms of the dual photon, the other 
in terms of the Wilson line. The former description is valid for 
$\hgamma\gg 1$, where it results in a two-dimensional 
sigma-model with target space given by the three-dimensional Coulomb 
branch.
As shown in Appendix (based on \cite{nd}), 
this Coulomb branch is equivalent to
the Higgs branch of Model A as a toric manifold and they have the same
Kahler class. We stress that, at this juncture,
we have made neither assumption nor conjecture about the 
properties of the three-dimensional theory. For the other description, 
in terms of the Wilson line, we may make progress in the regime 
$\hgamma\ll 1$ where the tower of Kaluza-Klein modes may be integrated out 
as in the previous section, resulting in the LG-model \eqn{Wtilde}.
The parameter which interpolates between the two
regimes, $\hat{\gamma}$, is a D-term deformation, and therefore
the holomorphic data --- the Kahler class of the sigma model and 
the twisted superpotential of the LG model --- can be extrapolated to 
all values of $\hat{\gamma}$.
Thus, under the assumption of the absence of
the phase transition between the two regimes
$\hgamma\gg 1$ and $\hgamma\ll 1$,
one can rederive the two-dimensional mirror symmetry \cite{HV}
from what we know for sure about three-dimensions
(up to the treatment of the divergent term in
(\ref{infpro}) that does rely on some asumption
on three-dimensional theories, to be discussed in
Appendix \ref{app:A}).
From this perspective, we see that two-dimensional mirror symmetry
is indeed an exchange between the dual photon and Wilson line description,
as in the free theory in Section \ref{subsec:ab}.

In other words, given the two-dimensional mirror symmetry
established in \cite{HV},
this observation suggests that there is
no phase transition between $\hgamma\gg 1$ and $\hgamma\ll 1$.
Semi-classical analysis of the two-dimensional theory
implies \cite{HV} that
the Kahler metric is the flat one in the LG mirror of the toric
sigma model (that arizes in $\hgamma\gg 1$ in the present set-up).
If there is no phase transition, the Kahler metric for $\Theta_p$
in the regime $\hgamma\ll 1$
should not be too different from the flat metric.

\subsection{The D-term at $\hgamma\ll 1$}

\newcommand{\lTheta}{\mbox{\small $\mit\Theta$}}

In fact, in the regime $\hgamma\ll 1$ one can
analyze also the D-term since we have small expansion parameters
$\hgamma/\Theta^2$ and $\hgamma$ as noted in the previous section.
The integration over the matter (including all the KK modes)
can be performed exactly at the one-loop level.
This leads to the following
effective metric on the Coulomb branch
\beq
\dd \widehat{s}^2
=\sum_{p,q}\left({\delta_{p,q}\over \widehat{e}_p^2}
+{1\over 2}
\sum_{i,n}{\hQ_i^p\hQ_i^q\over |\widehat{\sigma}_i+in/R|^2}
\right)\dd \overline{\widehat{\sigma}}_p\dd\widehat{\sigma}_q,
\label{dsh}
\eeq
where $\widehat{\sigma}_i=\hQ_i^p\widehat{\sigma}_p+\hm_i$.
The matter integral also induces other interactions of
vector multiplet fields including Kaluza-Klein modes
$\widehat{\Sigma}(n)$. Each of them is a power of
$\widehat{\Sigma}(n)$'s with some derivatives, divided by some power of
$|\widehat{\sigma}_i|$'s (of at least second order).
Thus, (\ref{dsh}) is the leading correction
in the $\hgamma/\Theta^2$ and $\hgamma$ expansion,
even after integrating out the vector multiplet KK
modes and higher frequency modes of $\widehat{\Sigma}$.

The infinite sum in (\ref{dsh}) can be perfomed by using the formula
\be
\sum_{n=-\infty}^{\infty}\frac{1}{x^2+(a+2\pi n)^2}
=\frac{\sinh x}{2x(\cosh x-\cos a)}
\ee
 which is $\sim 1/2x$ for large $x$.
Noting the rescaling $\widehat{\sigma}_p=\lTheta_p/2\pi R$ (\ref{resc})
and the $R$ dependence (\ref{mi}) of $\hm_i$,
we find that the Kahler metric behaves as
\beq
\dd \widehat{s}^2=\sum_{p,q}\left({\delta_{p,q}\over \hgamma_p}
+{1\over 4}\sum_{i=1}^N
{\hQ_i^p\hQ_i^q\over (2\pi R \hm_i+{\rm Re}\hQ_i^r\lTheta_r)}
\right)\dd\overline{\lTheta}_p\dd\lTheta_q.
\eeq
If we introduce the notation, $Y_i=\hQ_i^p\Theta_p+r_i$
which obeys $\sum_{i=1}^NQ_i^aY_i=r^a$ and appears in the superpotential as
$\widetilde{W}=-\Lambda\sum_{i=1}^N\e^{-Y_i}$, the metric is written as
\beq
\dd \widehat{s}^2=\sum_{p=1}^{N-k}{1\over \hgamma_p}|\dd\lTheta_p|^2
+{1\over 4}\sum_{i=1}^N{|\dd y_i|^2\over \log(\Lambda_{\rm UV}/\Lambda)
+{\rm Re}\,y_i}.
\label{dsh2}
\eeq
We see that it approaches the flat metric 
$\sum_{p=1}^{N-k}|\dd\lTheta_p|^2/\hgamma_p$
in the two-dimensional continuum limit
$\Lambda_{\rm UV}/\Lambda\to\infty$.

If we were allowed to take
the limit $\Lambda_{\rm UV}/\Lambda\to \infty$
first,
then the Kahler potential would stay flat even in the regime
$\hgamma\gg 1$ (with a vanishingly small coefficient at
$\hgamma\to \infty$ as in \cite{HV}).
In the next subsection, instead of discussing whether this exchange of
limits is valid, we turn to the mirror of
the Model B compactification with $\hgamma \ll 1$.

\subsection{3d Mirror Symmetry beyond IR duality}

As mentioned before, three-dimensional mirror symmetry is
originally conjectured as an infra-red duality.
However, there is a proposal that in 
fact it extends to all energy scales, with a mild modification
of the field contents and the interaction \cite{ks}.
Let us introduce

\noindent
{\bf Model A$^{\prime}$}: $U(1)^{2N-k}=\prod_{a=1}^kU(1)_a\times
\prod_{p'=1}^{N-k}U(1)'_{p'}\times\prod_{\widetilde{p}=1}^{N-k}
\widetilde{U}(1)_{\widetilde{p}}$ gauge theory with $N$ chiral multiplets
$\Phi_i$ of charge $Q_i^a$ and $R_{ip'}$ under $U(1)_a$ and
$U(1)'_{p'}$ but neutral under $\widetilde{U}(1)_{\widetilde{p}}$.
Non-zero Chern-Simons couplings are
$k^{ab}={1\over 2}\sum_{i=1}^NQ_i^aQ_i^b$,
$k_{p'q'}={1\over 2}\sum_{i=1}^NR_{ip'}R_{iq'}$,
$k^a_{p'}=\sum_{i=1}^NQ_i^aR_{ip'}$ and
$k_{p'}^{\widetilde{q}}=\delta_{p'}^{\widetilde{q}}$.
The theory is parametrized by the gauge coupling constants
$e_a,e_{p'},\widetilde{e}_{\widetilde{p}}$,
and $N$ combinations of the FI parameters $\zeta^a$, $\zeta'_{p'}$ and
the real masses $m_i$ (the FI parameters of
$\widetilde{U}(1)_{\widetilde{p}}$ can always
be set equal to zero by a field redefinition).

\noindent
Here we define $R_{ip'}$ as solutions to
\beq
\sum_{i=1}^N\hQ_i^pR_{iq'}=\delta^{p}_{q'},
\eeq
(the ambiguity of shift by $Q_i^a$ actually does not matter).
Model A${}'$ may be derived by the argument of \cite{Dave}, starting with
the all scale ${\cal N}=4$ mirror symmetry proposal of \cite{ks}.
According to this proposal,
Model A${}'$ in the limit $e_a,e'_{p'}\to\infty$ is exactly dual
to Model B, provided $\widetilde{e}_p=\widehat{e}_p$
and the FI and masses are related by (\ref{mm}) with
$\zeta'_{p'}-{1\over 2}\sum_{i=1}^NR_{ip'}m_i=\sum_{i=1}^NR_{ip'}\hm_i$.

The action is quadratic in $\widetilde{V}_{\widetilde{q}}$
and one may dualize it to a periodic chiral superfield
$P_{\widetilde{q}}$.
Then, the Higgs branch is of $N+(N-k)-[k+(N-k)]=N-k$ 
dimensions, as in the case of Model A.
In fact, it is identical to the Higgs branch of Model A
as a toric manifold with a Kahler class.
The difference is that now the torus fibres have a constant size
deep in the interior of the base of the fibration.
For example, let us consider the case of $N=2$, $k=1$
with $Q_i=1$, $Q'_1=1/2,Q'_2=-1/2$. Then, the metric
of the Higgs branch is given by
\beq
\dd s^2
=\left(\frac{1}{\widetilde{e}^2}
+\frac{\zeta}{\zeta^2-\rho^2}\right)\dd\rho^2
+\left(\frac{1}{\widetilde{e}^2}
+\frac{\zeta}{\zeta^2-\rho^2}\right)^{-1}
\dd\varphi^2,
\eeq
where $\rho$ is the coordinate of the
base $[-\zeta,\zeta]$ and $\varphi$ is the coordinate of period $2\pi$
of the torus fibre.
We see that deep in the interior $-\zeta\ll \rho \ll \zeta$ of the base,
the torus fibre has a constant radius. If $\zeta$
is much larger than $\widetilde{e}^2$, the radius is approximately
$\widetilde{e}$.

After compactification,
we obtain a non-linear sigma model on this squashed toric manifold
(which we shall call squashed toric sigma model).
We give $R$ dependence to $2\pi R\zeta$ as (\ref{ascale}), and it is much
larger than $\hgamma=2\pi R\widetilde{e}^2$ which is held fixed.
Thus, the squared radius of the torus fibre is
$\hgamma_p=2\pi R \widetilde{e}_p^2$ at high enough energies
in the two-dimensional theory.
Since large and small $\hgamma$ are different simply
by the squashing factor, it is hard to imagine that there is a phase
transition between $\hgamma\gg 1$ and $\hgamma\ll 1$.

The squashed toric sigma model is in the class of theories studied recently
in \cite{HK} in detail.
As in \cite{HK}, one can apply the argument of mirror symmetry
in \cite{HV}.
The dual of $\Phi_i$ and $P_{q}$
are twisted chiral superfields $Y_i$ and $\widetilde{Y}_q$ of period
$2\pi i$ ($\widetilde{Y}_q$ is in fact the fieldstrength
$2\pi R \widetilde{\Sigma}_q$ of $\widetilde{U}(1)_q$).
The twisted superpotential is
\beq
\widetilde{W}=\sum_{a=1}^k\Sigma_a(Q_i^aY_i-r^a)
+\sum_{q=1}^{N-k}{\Sigma'}^q(R_{iq}Y_i-\widetilde{Y}_q)
+\sum_{i=1}^N\e^{-Y_i}.
\eeq
The semi-classical Kahler metric is given by
\beq
\dd s^2=\sum_{q=1}^{N-k}{1\over \hgamma_q}|\dd \widetilde{y}_q|^2
+{1\over 2}\sum_{i=1}^N{|\dd y_i|^2\over \log(\Lambda_{\rm UV}/\Lambda)
+{\rm Re}\,y_i}.
\label{ds3}
\eeq
After integrating out $\Sigma_a$ and ${\Sigma'}^q$,
we obtain the constraints $\sum_{i=1}^NQ_i^aY_i=r^a$, together with 
$\sum_{i=1}^NR_{iq}Y_i=\widetilde{Y}_q$, which are solved by
$Y_i=\sum_{q=1}^{N-k}\hQ_i^p\widetilde{Y}_q+r_i$.
(It has been conjectured by Fendley and Intriligator
that the supersymmetric
squashed $S^2$ sigma model, the supersymmetric version of the
sausage model of \cite{sausage}, is mirror to sine-Gordon model
of finite Kahler potential.)

We see that the compactification of Model B is essentially the same
as this two-dimensional mirror of the squashed toric sigma model.
It has the same superpotential $\sum_{i=1}^N\e^{-Y_i}$ and
the behaviour of the Kahler metric (\ref{dsh2}) is similar to
the semi-classical metric (\ref{ds3}) of the 2d mirror.\footnote{
It is curious to note that the second terms in (\ref{dsh2})
and (\ref{ds3}), which are both vanishing in the two-dimensional
continuum limit, differs only by a factor of $2$.
It would be interesting to understand why the behaviour is similar
and why they differ by the factor of $2$.}
Thus, the all scale mirror symmetry conjecture \cite{ks}
is consistent with the
$1+1$ dimensional mirror symmetry that is already established.

\section{Vortex-Electron Exchange}
\label{sec:vor}

\newcommand{\tX}{\tilde{X}}

In \cite{ahiss} it was argued that mirror symmetry of 
three-dimensional gauge theories can be 
interpreted as an exchange of the electron and vortex 
descriptions (see also \cite{ks}). 
More precisely, mirror symmetry exchanges 
Nielson-Oleson vortices on the Higgs branch with bound states 
of logarithmically confined electrons on the Coulomb 
branch. In this section we will elaborate on this interpretation in 
our models, and study its relation to the compactification
analysis.

As usual, the mass of a particle is bounded from below by the central charge
of the supersymmetry algebra.
${\cal N}=2$ supersymmetry algebra contains one real central charge
which is a linear combination of conserved charges.
In Model A, on the Higgs branch, there are $k$ topological charges ---
 the vortex numbers.
Note that there are no Noether charges since the global symmetry
$U(1)^{N-k}$ is
generically spontaneously broken.
In contrast, in Model B on the Coulomb branch, 
the global symmetry $U(1)^k$ is unbroken
and there are $k$ Noether charges.
Thus, in both theories, the ${\cal N}=2$ central charge
is a linear combination of $k$ integers.

Consider firstly the Coulomb branch of Model B, as described 
in the Appendix. The possible BPS states of the theory, which 
lie in short representations of the supersymmetry algebra, are 
associated to (suitably ordered) products of chiral operators,  
\be
X(n_i)=\prod_{i=1}^N\widehat{\Phi}_i^{n_i}
\ee
i.e. $n_i\geq 0$ electrons of the $i^{\rm th}$ type. However, 
in three-dimensional gauge theories with massless gauge bosons,
the $1/r$ fall-off of electric fields ensures 
that any state charged under a local current has logarithmically 
divergent mass. Thus, on the Coulomb branch,
all finite mass states in the theory are 
associated to gauge invariant operators. The operator $X(n_i)$ 
is gauge invariant if and only if the positive integers $n_i$ 
satisfy,
\be
\sum_{i=1}^Nn_i\hQ_i^p=0\ \ \ \ \ \ \ \ \forall\ p=1,\cdots,N-k
\ee
The solutions to this equation, if they exist, are given 
by $n_i=\sum_{a=1}^kp_aQ_i^a$ 
for some integers $p_1,\ldots, p_k$.
The integers $p_a$ can be identified as the
Noether charge of $X(n_i)$ associated with the $U(1)^k$ global
symmetry group.\footnote{
The global symmetry $U(1)^k$ acts on $\hPhi_i$ with charge
the $\hR_{ia}$ complementary to $\hQ_i^p$.
If $\hR_{ia}$ are chosen to obey $\sum_{i=1}^NQ_i^a\hR_{ib}=\delta^a_b$,
the charge of $X(n_i)$ is
$\sum_{i=1}^Nn_i\hR_{ia}=\sum_{i,b}p_bQ_i^b\hR_{ia}=p_a$.}
Thus, if there exists a $k$-vector charge $p_a$ such 
that the resulting $n_i$ are all non-negative, then there exists a 
corresponding BPS bound state of $\sum_{i=1}^Nn_i$
logarithmically confined electrons. In this case the mass of the 
state is determined by the central charge, which is not renormalized, 
and is given by,
\be
M(n_i)=\sum_{i=1}^Nn_i\hat{m}_i
=\sum_{a=1}^k\left(\sum_{i=1}^NQ_i^a\hm_i\right)p_a.
\label{cmass}
\ee
The gauge invariance of this state ensures that the mass 
is independent of $\hat{\phi}_p$,
the position on the Coulomb branch.
Indeed the sum of the effective mass
$\hM_i=\hQ_i^p\hat{\phi}_p+\hm_i$
of the constituents
is the same as the central charge (\ref{cmass}),
$\sum_{i=1}^Nn_i\hM_i
=\sum_{i=1}^Nn_i\hm_i=M(n_i)$.

For a charge $p_a$ such that some of the resulting $n_i$ are negative,
one cannot find a gauge invariant chiral operator as a combination of
positive powers of $\hPhi_i$'s.
Instead, there exists an operator that involves  
both chiral and anti-chiral superfields;
\beq
\tX(n_i)=\prod_{i\in I}\widehat{\Phi}_i^{n_i}
\prod_{j\notin I}(\widehat{\Phi}^\dagger_j)^{-n_j}.
\eeq
where
\beq
I:=\{\,i\,;\,n_i\geq 0\}.
\eeq
This is gauge invariant (in the Wess-Zumino gauge) and has the right
charge $p_a$ under $U(1)^k$ global symmetry group.
The mass of the state associated to $\tX(n_i)$ is subject to renormalization.
In fact, the sum of the effective masses 
$\sum_{i=1}^N|n_i|\hM_i$ depends on the values of $\hphi_p$
and is in general larger than the central charge (\ref{cmass}).
Nevertheless, if we restrict to the locus on the Coulomb branch,
\be
\sum_{p=1}^{N-k}Q_i^p\hat{\phi}_p+\widehat{m}_i=0\ \ \  \ \ \ \ \ \ 
\forall\ i\notin I.
\label{cloci}
\ee
then the effective mass of the electrons $\widehat{\Phi}_i$ ($i\notin I$)
is vanishing and the sum $\sum_{i=1}^N|n_i|\hM_i$
becomes $\sum_{i=1}^Nn_i\hM_i=M(n_i)$ and classically saturates the BPS bound.
However, there is still room for renormalization and the state 
associated to $\tX(n_i)$ does not lie in a short multiplet.

Let us now turn to the Higgs branch of Model A where the particle states 
are Nielsen-Olesen vortices. We consider a vortex arising from 
non-trivial winding of the $U(1)$ subgroup
of the gauge group $U(1)^k$, defined by
$\e^{i\theta}\hookrightarrow (\e^{ip_a\theta})$,
 under which the chiral multiplet $\Phi_i$ has 
charge $n_i=\sum_{a=1}^kp_aQ_i^a$. Note that we have employed the same 
notation, $p_a$ and $n_i$, as used in the discussion of the Coulomb branch 
and, indeed, these quantities will be identified with each other under 
mirror symmetry. Let us examine under what circumstances the vortices 
saturates the classical BPS bound.
Specifically, consider a vortex with unit flux 
${-1\over 2\pi}\int \dd v=1$. While such vortices exist at all points 
of the Higgs branch, they saturate the classical BPS bound
only at specific points. 
To see this, we need the result that the Bogomoln'yi equations can only be 
satisified if \cite{witten}
\be
\Phi_i=0 \ \ \ \  \ \ \ \ \ \forall\ i\notin I
\ee
This requirement is equivalent to the statement that a line bundle of 
negative degree has no non-zero holomorphic section. Notice that this 
restriction to loci of the Higgs branch is mapped under mirror symmetry 
to \eqn{cloci}.

Now let us ask under what circumstance the vortices lie in 
short representations of the supersymmetry algebra. To see this 
we must quantize the zero-modes of the vortex. 
The vortex system has $\sum_{i\in I}n_i$ bosonic 
zero modes \cite{morples}\footnote{To see this, one must adapt the results of 
\cite{morples} to {\em based} vortices which asymptote to the vacuum at a 
given point on the ${\bf P}^1$ worldsheet.}.
There are also fermionic zero modes.
In three-dimensions each chiral multiplet, $\Phi_i$, contains 
a single Dirac fermion $\psi_i$. For $i\in I$, each $\psi_i$ 
has $n_i$ zero modes while the conjugate spinor $\bar{\psi}_i$ 
has none\footnote{Upon dimensional reduction to two dimensions only, 
say, left moving spinors have zero modes while right-movers have 
none.}. These are paired by the unbroken supersymmetry of the vortex  
with the bosonic zero modes. However, fermionic zero modes arise also 
from $\Phi_i$ with $i\notin I$. In this case $\psi_i$ has no 
zero modes, while $\bar{\psi}_i$ has $-n_i$, which are not paired 
with bosonic zero modes. Thus the total 
number of fermionic zero modes is given by $\sum_{i=1}^N|n_i|$.
This mismatch of bosonic and fermionic zero modes
reflects the fact that the vortices lie in 
long multiplets unless $n_i\geq 0$ for all $i$, in agreement
with the situation on the Coulomb branch.
Note that when this occurs, the mass of the BPS vortex is given by
\be
M(n_i)=\sum_{i=1}^k p_a\zeta_a
\ee
which agrees with \eqn{cmass} under the identification of the 
mirror map (note that we have set $m_i=0$ for Model A).

We note that the number $\sum_{i\in I}n_i$ of bosonic zero modes
of a Model A vortex agrees with the number of holomorphic constituent
components of the logarithmically confined states
on the Model B Coulomb branch.
This might suggest that the electron $\hPhi_i$ of Model B
can be interpreted as the fractional vortex which has winding number one
in the $\Phi_i$ component.
This picture is in fact consistent with
the generation of the superpotential \eqn{Wtilde}
in the compactified theory, in comparison with the two-dimensional
derivation \cite{HV}.
We have seen that each term $\e^{-\hQ_i^p\Theta_p-r_i}$
is generated by the loop of $\hPhi_i$ Kaluza-Klein modes.
On the other hand, in two-dimensions,
the same term is generated by a one $\Phi_i$ vortex-instanton
in the theory with an extended gauge symmetry \cite{HV}.
It would also be interesting to analyze
the fractional instanton effect, directly
in the theory without extended gauge symmetry.

\section{Concluding Remarks}
\label{conclude}

In this section we would like to briefly summarize the 
calculations performed in this paper, and suggest some 
avenues for further research. 

The primary accomplishment was to compactify 
$2+1$-dimensional Abelian-Chern-Simons mirror pairs with four 
supercharges on a circle of radius $R$. We showed that in order 
to arrive at a non-trivial interacting theory in the continuum limit 
$R\rightarrow 0$, both the parameters and the fields  
of the three-dimensional theory must scale with $R$. When this scaling 
is performed, the theory on the Higgs branch, denoted as Model A 
in the text, reduces to a $1+1$-dimensional non-linear sigma model 
with a toric target space. The tower of KK modes play no role for 
this theory, decoupling as $R\rightarrow 0$. In contrast, for the 
theory on the Coulomb branch, denoted as Model B, the necessary 
scalings ensure that the KK modes do not decouple. Rather, each 
contributes to the superpotential and the total may be resummed, 
resulting in the exact superpotential \eqn{goodinnit}. In the 
limit $R\rightarrow 0$, this reduces to the Toda-like LG 
superpotential, which was shown in \cite{HV} to be dual to the 
toric sigma-model.

It is noteworthy that the resummation of loop effects of
the infinite tower of KK modes captures non-perturbative effects 
in $1+1$ dimensions. This phenomenon is not new: it has 
occured previously in compactification from $3+1$ to
$2+1$ dimensions \cite{SandS}, as well as in compactification from 
$4+1$ to $3+1$ dimensions \cite{Nikita}.

As mentioned in the introduction, the mirror transformation in 
three dimensions may be thought of as a generalization of 
scalar-vector duality. In particular, the variable parameterizing 
the Higgs branch of Model A is mapped to the dual photon of 
Model B. However, when considering the compactification of Model 
B, we worked not with the dual photon, but rather with the original 
gauge field, in the guise of a Wilson line, and the resulting 
LG theory is a description in 
terms of these variables. As discussed in Section \ref{subsec:ab}, 
the scalar-vector duality of free-Abelian theories in  
three-dimension reduces upon dimensional reduction to scalar-scalar 
duality in two-dimensions. One of the main messages of this paper is 
that, at least for the class of models considered, this correspondence 
continues to hold even in the interacting case. 

In Section 4, we analyzed in more detail the dependance of the 
compactification on coupling constants of Model A and Model B and, 
in particular, on the dimensionless ratios, $\gamma=e^2_{\rm 3d}R$ and 
$\widehat{\gamma}=\widehat{e}^2_{3d}R$. We noted that the derivation 
given in Section 3 was valid only in the regime 
$\widehat{\gamma}\ll 1$, while the use of three-dimensional mirror 
symmetry as an infra-red duality would require $\widehat{\gamma}\gg 1$. 
We interpreted the fact that we obtained the correct answer as 
evidence for the lack of a phase transition as $\widehat{\gamma}$ 
is varied. Moreover, we repeated our discussion for the all-scale 
three-dimensional mirror pairs conjectured in \cite{ks}. To extend 
mirror symmetry beyond the infra-red limit requires a modification 
of the theory on the Higgs branch; this was denoted Model A$^\prime$ in 
the text. As a test of the conjecture that these theories are indeed 
mirror on all length scales, we examined the 
Kahler potential, a quantity not protected by supersymmetry. We found that 
a finite Kahler potential for the LG theory in $1+1$ dimensions   
corresponds to a squashing of the metric of the toric sigma-model, 
in agreement with expectations \cite{HK}.

\subsection{Future directions}

There remain several open questions, some of which we list here. 
Firstly, we considered only Abelian gauge theories in this paper. 
In $1+1$ dimensions, the derivation of mirror symmetry given in 
\cite{HV} does not straightforwardly apply to non-Abelian theories. 
Moreover, the dual of non-linear sigma models arising from 
non-Abelian gauge theories (such as Grassmannians \cite{grass})
are unknown, although 
there are indications of the existence of LG type mirrors \cite{EHX}. 
In contrast, there exist many three-dimensional non-Abelian mirror 
pairs. One may therefore use the techniques presented here in order 
to find new non-Abelian mirror pairs in two dimensions. 

Secondly, an extremely interesting class of mirror pairs in 
$1+1$ dimensions are the $(2,2)$ superconformal theories with 
compact Calabi-Yau target spaces. For reviews and references see 
\cite{brian,HV,mor}. It would be interesting if this could also be 
related three-dimensional theories.

Finally, as shown in \cite{Dave} (and also in Appendix~\ref{app:A}),
our three-dimensional ${\cal N}=2$ mirror theories can be obtained
from the RG flow of ${\cal N}=4$ mirror theories perturbed by
an operator which  partially breaks supersymmetry. In other words, 
the route we have taken to arrive at $1+1$ dimensional mirrors is: 
3d ${\cal N}=4$ $\Rightarrow$3d ${\cal N}=2$ $\Rightarrow$ 2d $(2,2)$.
It is natural to ask\footnote{We thank C.Vafa for 
doing so.} if one may exchange the order of partial supersymmetry
breaking and compactification: 
3d ${\cal N}=4$ $\Rightarrow$
2d $(4,4)$ $\Rightarrow$ 2d $(2,2)$. Along this route, one 
first compactifies the ${\cal N}=4$ mirror pairs to obtain
duality between two-dimensional $(4,4)$ theories \cite{ds}, before 
deforming the resulting pairs by an operator that breaks $(4,4)$
supersymmetry to $(2,2)$.

\section*{Acknowledgements}

We would like to thank Bobby Acharya, Ted Baltz, J.~de~Boer,
Nick Dorey, Richard Easther, P.~Fendley,
Brian Greene, K.~Intriligator, Dan Kabat, A.~Kapustin,
K.~Ohta, S.~Sethi, Matt Strassler, C.~Vafa and T.~Yokono
for useful discussion.
MA, KH and AK thank Aspen Center for Physics
and ITP at UC Santa Barbara,
where parts of this work were done.
DT is grateful to Center for Theoretical Physics, MIT and to the
Harvard Theory group for hospitality. This work was
supported by the US Department of Energy.
MA, KH and AK are supported in part by NSF-PHY-9802709, NSF-DMS 9709694,
and the U.S. Department of Energy under contract \#
DE-FC02-94ER40818, respectively, and also by NSF-PHY-9907949.

\appendix{Regularization and the Mirror Map}
\label{app:A}

In this appendix,
we describe the regularization scheme of Model A and B which
has been used in the text so far.
In particular, we derive the mirror map (\ref{mm}) and also
explain why the second term in (\ref{infpro})
\beq
-\sum_{i=1}^N\hQ_i^p\log\left[{1\over 2\pi R}\prod_{n=1}^{\infty}
\left({n^2\over R^2}\right)\right]
\eeq
can be set equal to zero.
We will also show that the final result (\ref{Wtilde})
is essentially independent of the choice of the regularization scheme.

\subsection{Deformation of finite ${\cal N}=4$ mirror pairs}

Model A is not finite unless
$\sum_{i=1}^NQ_i^a=0$ for all $a$ (and similarly for Model B);
The one-point function of
the auxiliary field
\beq
\langle D_a\rangle\propto \sum_{i=1}^NQ_i^a\int {\dd^3k\over k^2+\cdots},
\label{div}
\eeq
is linearly divergent.
Thus, both Model A and Model B have to be regularized and
the FI parameters must be renormalized.
We will show that
the finite ${\cal N}=4$ mirror pairs
can be used as the cut-off theories.


We consider the following
${\cal N}=4$ pair from \cite{dhooy}

\noindent
{\bf Model A${}_{(4)}$:}\ $U(1)^k$ gauge theory of
vector multiplets $(V_a,\Psi_a)$ with
$N$ hypermultiplets $H_i=(\Phi_i,\Phi_i^{\vee})$
of charge $Q_i^a$.
Only the real FI parameters $\zeta^a_{(4)}$ and the real masses
$m^{(4)}_i$ for $H_i$ are turned on.

\noindent
{\bf Model B${}_{(4)}$:}\
$U(1)^{N-k}$ gauge theory of $(\widehat{V}_p,\widehat{\Psi}_p)$
 with $N$ hypermultiplets
$\widehat{H}_i=(\hPhi_i,\hPhi_i^{\vee})$ of charge $\hQ_i^p$.
The real FI parameters $\hz_{(4)}^{p}$ and the real masses $\hm_i^{(4)}$ 
for $\widehat{H}_i$ are turned on, but the complex FI/masses are
turned off.

\noindent
The parameters are related by the ${\cal N}=4$ mirror map
\beq
\zeta_{(4)}^a=\sum_{i=1}^NQ_i^a\hm_i^{(4)},~~~
-\sum_{i=1}^N\hQ_i^pm_i^{(4)}=\hz_{(4)}^p.
\label{mm4}
\eeq

As in \cite{Dave}, we give a background value $X$
to the scalar component of the vector multiplet for
a $U(1)$ subgroup of the R-symmetry.
This gives a mass $2X$ to $\Psi_a$ and $\widehat{\Psi}_p$,
and the masses of the hypermultiplet fields are changed to
\beqa
&&m^2(\Phi_i)=|Q_i^a\Psi_a|^2+(Q_i^a\phi_a+m_i^{(4)}+X)^2,\\
&&m^2(\Phi_i^{\vee})=|Q_i^a\Psi_a|^2+(-Q_i^a\phi_a-m_i^{(4)}+X)^2,\\
&&m^2(\widehat{\Phi}_i)=
|\hQ_i^p\widehat{\Psi}_p|^2
+(\hQ_i^p\widehat{\phi}_p+\hm_i^{(4)}-X)^2,\\
&&m^2(\widehat{\Phi}_i^{\vee})=
|\hQ_i^p\widehat{\Psi}_p|^2
+(-\hQ_i^p\widehat{\phi}_p-\hm_i^{(4)}-X)^2.
\eeqa
We will send $X$ to infinity, keeeping fixed
the following mass parameters
\beq
m_i=m_i^{(4)}+X,~~~
\hm_i=\hm_i^{(4)}-X.
\label{para}
\eeq
Then, $\Phi_i$, $\widehat{\Phi}_i$ have finite masses but
$\Psi_a$, $\widehat{\Psi}_p$, $\Phi_i^{\vee}$ and
$\widehat{\Phi}_i^{\vee}$ have masses that diverges as $\sim 2X$.
Thus, it is appropriate to integrate out these heavy fields.
Integration of the neutral fields
$\Psi_a$, $\widehat{\Psi}_p$ simply sets these fields
to zero.
Integration of the charged fields generates Chern-Simons and FI terms
for the ${\cal N}=2$ vector multiplets.
To find the precise form,
we compactify the theory on the circle of radius $R$
where the three-dimensional theory
is recovered by taking the decompactification limit $R\to\infty$.
This is simply for a technical reason;
one can use the machinery of computation
developed in Section \ref{subsec:cptB}.

As in Section \ref{subsec:cptB},
the generated superpotential $\sDelta\widetilde{W}$
in Model A${}_{(4)}$ is
\beq
{\partial\sDelta\widetilde{W}\over \partial
\Sigma_a}=\sum_{i=1}^NQ_i^a\log(\e^{\pi R\Sigma_i}-\e^{-\pi R\Sigma_i})
+\sum_{i=1}^NQ_i^a\log Z,
\label{dse}
\eeq
where $\Sigma_i=-Q_i^a\Sigma_a-m_i+2X$ and $Z$ is the infinite product
\beq
Z={1\over 2\pi R}\prod_{n=1}^{\infty}{n^2\over R^2},
\label{defZ}
\eeq
that appears also in (\ref{infpro}).
Since $\Sigma_i$ is large in the limit $X\to +\infty$,
$\e^{\pi R\Sigma_i}$ is dominant in the log
of the first term in (\ref{dse}),
 compared to
$\e^{-\pi R\Sigma_i}$ that vanishes
in the decompactification limit $R\to\infty$.
Thus, (\ref{dse}) is essentially
$\sum_{i=1}^NQ_i^a(\pi R\Sigma_i+\log Z)$.
By integration, we find
\beq
\sDelta\widetilde{W}=
-{\pi R\over 2}\sum_{i,a,b}Q_i^aQ_i^b\Sigma_a\Sigma_b
-2\pi R \sum_a\left({1\over 2}\sum_{i}Q_i^am_i
-\sum_{i}Q_i^a(X+{1\over 2\pi R}\log Z)\right)\Sigma_a.
\eeq
Comparing with (\ref{whassupnow}) and (\ref{WHASSSUP}),
we find that the CS and FI couplings are given by
\beqa
&&k^{ab}={1\over 2}\sum_{i=1}^NQ_i^aQ_i^b,
\label{Az1}\\
&&\zeta^a=\zeta_{(4)}^a+{1\over 2}\sum_{i=1}^NQ_i^am_i
-\sum_{i=1}^NQ_i^a\left(X+{1\over 2\pi R}\log Z\right)
\label{Am1}
\eeqa
The same procedure gives the CS and FI coupling of Model B;
\beqa
&&\widehat{k}^{pq}=-{1\over 2}\sum_{i=1}^N\hQ_i^p\hQ_i^q,\\
&&\hz^p=\hz_{(4)}^p-{1\over 2}\sum_{i=1}^N\hQ_i^p\hm_i
-\sum_{i=1}^N\hQ_i^p\left(X+{1\over 2\pi R}\log Z\right).
\label{Am2}
\eeqa
The change in the sign compared to (\ref{Az1})-(\ref{Am1}) is because
$\widehat{\Sigma}_i=-\hQ_i^p\widehat{\Sigma}_p-\hm_i-2X$
have large {\it negative} values.

\subsection{Zeta function regularization}

We notice that the last terms in (\ref{Am1}) and (\ref{Am2}) are divergent,
which is related to the divergence (\ref{div}).
At this point, we treat the infinite product
(\ref{defZ}) by zeta function regularization.
Defining $\zeta_R(s)=\sum_{n=1}^{\infty}(n/R)^{-s}=R^s\zeta(s)$
and noting that $\zeta(0)=-{1\over 2},
\zeta'(0)=-{1\over 2}\log(2\pi)$,
we find
\beq
Z={1\over 2\pi R}\e^{-2\zeta_R'(0)}
={1\over 2\pi R}\e^{-2\zeta'(0)-2\zeta(0)\log R}=1.
\label{zeta}
\eeq
Then, the $\log Z$ terms in (\ref{Am1}) and (\ref{Am2})  vanish.
Using the ${\cal N}=4$ mirror map (\ref{mm4}) and (\ref{para}),
we find that the $X$ linear terms also cancel out;
\beqa
&&\zeta^a=\sum_{i=1}^NQ_i^a\hm_i+{1\over 2}\sum_{i=1}^NQ_i^am_i,
\label{Amm1}\\
&&\hz^p=-\sum_{i=1}^N\hQ_i^pm_i
-{1\over 2}\sum_{i=1}^N\hQ_i^p\hm_i,
\label{Amm2}
\eeqa
In this way,
we recover the ${\cal N}=2$ mirror map (\ref{mm}).
We have seen that this mirror map is based on
the zeta function regularization (\ref{zeta}).
It is also clear that the second term of (\ref{infpro})
should be set equal to zero under this regularization scheme.

\newcommand{\onefigure}[2]{\begin{figure}[htbp]
         \caption{\small #2\label{#1}(#1)}
         \end{figure}}
\newcommand{\onefigurenocap}[1]{\begin{figure}[h]
         \begin{center}\leavevmode\epsfbox{#1.eps}\end{center}
         \end{figure}}
\renewcommand{\onefigure}[2]{\begin{figure}[htbp]
         \begin{center}\leavevmode\epsfbox{#1.eps}\end{center}
         \caption{\small #2\label{#1}}
         \end{figure}}

\begin{figure}[htb]
\begin{center}
\epsfxsize=4in\leavevmode\epsfbox{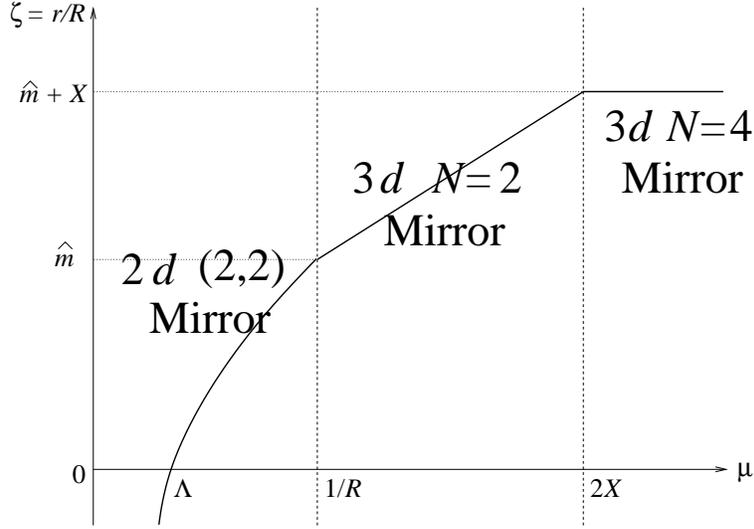}
\end{center}
\caption{Mirror symmetry at various scales.}
\label{mir}
\end{figure}

\subsection{Other regularization schemes}

We could have chosen another regularization scheme where
$\log Z$ does indeed diverge.
First, it is easy to see that the $\log Z$ term in
$\hz^p$ of (\ref{Am2}) cancels the second term of (\ref{infpro})
when we add $\widetilde{W}_{FI}+\widetilde{W}_{CS}$ to
$\sDelta\widetilde{W}$.
What about the $\log Z$ term in (\ref{Am1})?

To examine this, let us look at the effective FI coupling
on the Higgs branch of Model A (where we set $m_i=0$).
The third term in (\ref{Am1}) can be considered as
an expression of the one-loop integral of $\Phi_i^{\vee}$
\beqa
-\sum_{i=1}^NQ_i^a\left(X+{\log Z\over 2\pi R}\right)
&=&\sum_{i=1}^NQ_i^a{1\over 2\pi R}\sum_{n=-\infty}^{\infty}
\int {\dd^2k\over 2\pi}{1 \over k^2+{n^2\over R^2}+(2X)^2}
\nonumber\\
&=&\sum_{i=1}^NQ_i^a
\int {\dd^3k\over (2\pi)^2}{1\over k^2+(2X)^2}.
\eeqa
The effective FI coupling at energy $\mu$ is
obtained by restricting this integral in the range $|k|\geq \mu$
and adding the contribution from the $\Phi_i$ loop integral as well;
\beqa
\zeta^a(\mu)&=&\zeta^a_{(4)}+\sum_{i=1}^NQ_i^a
\left[\int_{|k|\geq\mu}{\dd^3k\over (2\pi)^2}{1\over k^2+(2X)^2}
-\int_{|k|\geq\mu}{\dd^3k\over (2\pi)^2}{1\over  k^2}\right]
\nonumber\\
&=&\left\{
\begin{array}{ll}
\zeta^a_{(4)}&\mu\gg 2X,\\
\zeta^a_{(4)}+\sum_{i=1}^NQ_i^a(\mu/\pi-X)&\mu\ll 2X.
\end{array}
\right.
\label{17199}
\eeqa
As noted before, $\zeta_{(4)}^a$ is chosen to be
$\zeta_{(4)}^a=\sum_{i=1}^NQ_i^a(\hm_i+X)$
and it cancels the $X$ linear term,
yielding
\beq
\zeta^a(\mu)=\sum_{i=1}^NQ_i^a\hm_i+\sum_{1=1}^NQ_i^a\mu/\pi,~~~
\mu\ll 2X.
\label{171}
\eeq
We find that $2X$
serves as the cut-off of Model A and the cut-off theory is provided by
the finite Model A${}_{(4)}$.

Upon compactification, the 2d sigma model is obtained by
renomalizing the FI parameter according to
the RG matching equation
\beq
\zeta^a(1/2\pi R)={1\over 2\pi R}
\left(\sum_{i=1}^NQ_i^a\log(1/2\pi R\Lambda)+r^a\right).
\eeq
Using (\ref{171}),
this dictates the $R$ dependence of $\hm_i$, as in (\ref{mi}).
The presence of the term $\sum_{i=1}^NQ_i^a\mu/\pi$ in $\zeta^a(\mu)$
only yields a shift of $r_i$ in (\ref{mi}) by $-1/\pi$.
This simply gives
a finite rescaling of the superpotential
$\widetilde{W}$
in the final result (\ref{Wtilde}).

The results of this appendix, and in particular the running of 
FI parameter, are summarized in Figure 1.

\appendix{The Coulomb Branch in Three Dimensions}
\label{app:B}


In this appendix we briefly review the results of \cite{nd}, explaining 
how the Coulomb branch of Model B comes about. We further determine the 
Kahler class of the moduli space and show that it recieves no 
quantum corrections beyond one-loop. 

To determine the Coulomb branch of Model B, we  begin with 
an examination of the classical potential energy,
\be
V=\sum_{p=1}^{N-k}\widehat{e}_p^2\left(\sum_{i=1}^N\hQ_i^p
|\widehat{\sPhi}_i|^2-\sum_{q=1}^{N-k}\widehat{k}^{pq}\hat{\phi}_q
-\hz^p\right)^2+\sum_{i=1}^N\hM_i^2|\widehat{\sPhi}_i|^2
\label{tangerine}\ee
Recall that $\hphi_p$ are the real scalars of the vector multiplets 
while $\widehat{\sPhi}_i$ are the complex scalars
of the chiral multiplets. The 
latter have real masses $\hM_i$ given by,
\be
\hM_i=\sum_{p=1}^{N-k}\hQ_i^p\hphi_p+\widehat{m}_i
\label{orange}\ee
Note in particular that the D-term (the first term in \eqn{tangerine}) 
includes the supersymmetric completion of the Chern-Simons coupling 
\cite{kimyeong} which, in our case, has coefficient
\be
\widehat{k}^{pq}=-\ft12\sum_{i=1}^N\hQ^p_i\hQ_i^q
\ee
At first glance, the appearance of this term in the potential seems to 
rule out the possibility of 
a Coulomb branch. Indeed, setting the chiral multiplets to zero, 
a  non-zero expectation value for $\hphi_p$ results 
in a non-zero energy,
\be
V=\sum_{p=1}^{N-k}\widehat{e}_p^2\left(\sum_{q=1}^{N-k}
\widehat{k}^{pq}\hat{\phi}_q+\hz^p\right)^2
\label{effpot}\ee
The fact that the FI parameters lift the Coulomb branch is well-known, 
while the fact that the supersymmetric completion of the Chern-Simons 
couplings perform a similar feat can easily be anticipated; 
the scalars $\hphi_p$ are the superpartners of the gauge fields, and the 
latter acquire gauge invariant masses from the Chern-Simons couplings. 

While there is no classical Coulomb branch, the siutation is quite 
different quantum mechanically, for both the FI and Chern-Simons 
parameters are renormalized upon integrating out the chiral multiplets. 
While perturbation theory is valid only in the regime 
$\hM_i\gg \widehat{e}_p^2$ for all $i,p$, the topological 
nature of both couplings\footnote{Note that the FI parameter may be 
thought of as a mixed Chern-Simons coupling between global and local 
currents and therefore enjoys the same one-loop non-renormalization theorem.} 
ensures that the one-loop result is exact \cite{coleman}, and given by,
\beqa
\widehat{k}^{pq}_{\rm eff}
&=&-\ft12\sum_{i=1}^N\hQ^p_i\hQ_i^q
+\ft12\sum_{i=1}^N\hQ_i^p\hQ_i^q\,{\rm sign}\,(\hM_i) \\ 
\hz^p_{\rm eff} &=& \hz^p+\ft12\sum_{i=1}^N\hQ^p_i\widehat{m}_i
\,{\rm sign}\,(\hM_i)
\nn\eeqa
The effective potential is then given by \eqn{effpot} with the 
parameters replaced by their quantum corrected 
versions. We see that the Chern-Simons coupling vanishes only in 
the regime,
\be
\hM_i\geq 0
\label{conditions}\ee
while the effective FI parameter also vanishes in this regime if the 
bare FI parameter is given by 
$\hz^p=\ft12\sum_i\hQ_i^p\widehat{m}_i$, in agreement with 
the mirror map \eqn{mm}. 

The $N$ inequalities \eqn{conditions}
define a region, $\Delta\subset{\bf R}^{N-k}$, 
parameterized by $\hphi_p$. 
$\Delta$ may or may not be compact depending on the charges $\hQ_i^p$. 
This defines ``half'' of the Coulomb branch. The other half comes from the 
dual photons, $\widehat{\varphi}^p$.
These parameterize a torus, ${\bf T}^{N-k}$, 
which is fibered over $\Delta$. Moreover, at the boundaries of $\Delta$, given 
by the equations $\hM_i=0$,  certain cycles of ${\bf T}^{N-k}$ degenerate. 
To see this, note that whenever $\hM_i=0$, we have integrated 
out a chiral multiplet of vanishing mass, and we therefore expect the 
linear combination of gauge fields under which $\widehat{\Phi}_i$ is charged 
to have a (possibly coordinate) singularity in the low-energy effective 
action. This in turn implies the degeneration of the cycle 
$\widehat{\varphi}^p\propto\hQ_i^p$.
One can also argue that this cycle 
vanishes using the symmetries of the dual photon \cite{ahiss,nd}.

Thus we arrive at the Coulomb branch of Model B; a torus 
${\bf T}^{N-k}$ fibered over a region $\Delta\subset {\bf R}^{N-k}$. 
In \cite{nd}, it was shown that the one-loop Coulomb branch coincides as  
a toric variety with the Higgs branch of Model A.
Moreover, the two moduli spaces have the same isometries, with the 
same fixed points, which is a statement that holds non-perturbatively.
We now determine the exact Kahler class of the 
Coulomb branch and show that it is not renormalized by quantum effects. 
To do this, we examine the 
expression for the supersymmetric low-energy effective action which, 
up to two derivatives, is given by
\be
{\cal L}=\int\dd^4\theta f(\hSlin_p).
\label{sslag}\ee
Here $\hSlin_p$ is the linear superfields
which contain the 
scalars $\hphi_p$ and the field strengths
$\widehat{v}_{p\mu\nu}$.
A linear superfield $\Slin$ has the following expansion 
(we set the fermionic components zero) 
\beqa
\Slin&=&\phi
+{1\over 2}\theta^+\btheta^-(D-iv_{01})
+{1\over 2}\theta^-\btheta^+(D+iv_{01})
\\
&&+{1\over 2}\theta^+\btheta^+(-v_{02}-v_{12})
+{1\over 2}\theta^-\btheta^-(v_{02}-v_{12})
+{1\over 4}\theta^+\theta^-\btheta^-\btheta^+(\partial_0^2-\partial_1^2
-\partial_2^2)\phi.
\nonumber
\eeqa
The crucial point here is that
the fieldstrength $v_{\mu\nu}$ appears only
in the $\theta\btheta$ terms whereas
$\phi$ appears in the $\theta^0$ and $\theta^2\btheta^2$ terms.
The bosonic part of the Lagrangian therefore reads
\beq
{\cal L}_b=g^{pq}\left(\partial_{\mu}\hphi_p\partial^{\mu}\hphi_q+
{1\over 2}\widehat{v}_{p\mu\nu}\widehat{v}_{q}^{\mu\nu}\right)
\eeq
where $g^{pq}:=\partial^2f/\partial\hphi_p\partial\hphi_q$.
After dualization we obtain
\be
{\cal L}_b=g^{pq}\partial_\mu\hphi_p\partial^\mu\hphi_q
+g_{pq}\partial_\mu\widehat{\varphi}^p\partial^\mu\widehat{\varphi}^q
\label{Lb}
\ee
where $g_{pq}$ is the inverse matrix of $g^{pq}$.
In particular, there are no cross-terms between $\hphi_p$ and 
$\widehat{\varphi}^p$.
This is unlike the effective action for ${\cal N}=4$ vector multiplet
which also contains an ${\cal N}=2$ chiral multiplet in a superpotential 
and does indeed lead to such cross-terms \cite{seiberg}.
 
The metric on the Coulomb branch (\ref{Lb}) can be written as
$\dd s^2=g_{pq}\dd z^p\dd\overline{z}^q$ where
\beq
z^p={\partial f\over \partial\hphi_p}+i\widehat{\varphi}^p,
\eeq
and this implies that $z^p$ are complex coordinates on the Coulomb branch. 
(This can also be proved by dualization on the superspace
\cite{HKLR}. See also Section 2.3 of \cite{DHO}.)
Thus, the Kahler form is given by
\beqa
\Omega&=&{i\over 2}g_{pq}\dd z^p\wedge\dd\overline{z}^q
=g_{pq}\dd\left({\partial f\over \partial\hphi_p}\right)
\wedge\dd\widehat{\varphi}^q
\nonumber\\
&=&\dd\hphi_p\wedge\dd\widehat{\varphi}^p.
\eeqa
The Kahler class is determined by measuring the area of 2-cycles
using $\Omega$.
A typical 2-cycle is
the circle fibration over a segment between two vertices of $\Delta$,
where the circle shrinks to zero size at the two ends.
Its $\Omega$ area is just $2\pi$ times the length of the segment,
which is determined exactly at the one-loop level.
Thus, the Kahler class is independent of the details of
the function $f$ and, in particular, is not corrected from the
one-loop result by further quantum effects.
Moreover, $\Omega$ coincides with the Kahler
form of the Higgs branch of Model A.


\begin{thebibliography}{99}

\small
\parskip=0pt plus 2pt

\bibitem{HV} K.~Hori and C.~Vafa,
``{\em Mirror Symmetry}'', hep-th/0002222.

\bibitem{IS}
K.~Intriligator and N.~Seiberg,
``{\em Mirror Symmetry in Three Dimensional 
Gauge Theories}'', Phys.Lett. {\bf B387} (1996) 513,
[hep-th/9607207]. 

\bibitem{stringy}
M.~Porrati and A.~Zaffaroni,
``{\em M-theory Origin of Mirror Symmetry in Three-Dimensional
Gauge Theories}'',
Nucl.\ Phys.\ B {\bf 490} (1997) 107
[hep-th/9611201]; \\
A.~Hanany and E.~Witten,
``{\em Type IIB Superstrings, BPS Monopoles,
and Three Dimensional Gauge  Dynamics}'',
Nucl.\ Phys.\ B {\bf 492} (1997) 152
[hep-th/9611230]; \\
K.~Hori, H.~Ooguri and C.~Vafa,
``{\em Non-Abelian Conifold Transitions and N = 4
Dualities in Three  Dimensions}'',
Nucl.\ Phys.\ B {\bf 504} (1997) 147
[hep-th/9705220].

\bibitem{ahiss}
O.~Aharony, A.~Hanany, K.~Intriligator, N.~Seiberg and M.~Strassler, 
``{\em Aspects of N=2 Supersymmetric Gauge Theories in Three Dimensions}'',
Nucl.Phys. {\bf B499} (1997) 67, [hep-th/9703110].

\bibitem{Dave} D. Tong, ``{\em Dynamics of N=2 Supersymmetric Chern-Simons Theories}'', 
JHEP {\bf 0007} (2000) 019, [hep-th/0005186]. 

\bibitem{nd} N. Dorey and D. Tong, ``{\em Mirror Symmetry and Toric Geometry in 
Three-Dimensional Gauge Theories}'', JHEP {\bf 0005} (2000) 018, [hep-th/9911094].

\bibitem{ks} A. Kapustin and M. Strassler,
``{\em On Mirror Symmetry in Three 
Dimensional Abelian Gauge Theories}'',
JHEP {\bf 9904} (1999) 021, [hep-th/9902033].

\bibitem{HK}
K.~Hori and A.~Kapustin,
``{\em Duality of the Fermionic 2d Black Hole and N=2 Liouville Theory
as Mirror Symmetry}'', hep-th/0104202.


\bibitem{ds} D-E. Diaconescu and N. Seiberg,
``{\em The Coulomb Branch of (4,4) 
Supersymmetric Field Theories in Two Dimensions}'',
JHEP {\bf 9707} (1997) 001, 
[hep-th/9707158].

\bibitem{ss} N. Seiberg and S. Sethi,
``{\em Comments on Neveu-Schwarz Five 
Branes}'', Adv. Theor. Math. Phys. {\bf 1} (1998) 259, [hep-th/9708085].

\bibitem{ak} M. Aganagic and A. Karch,
``{\em Calabi-Yau Mirror Symmetry as a 
Gauge Theory Duality}'', Class.Quant.Grav. {\bf 17} (2000) 919,
[hep-th/9910184].

\bibitem{dhooy}
J.~de Boer et al,
``{\em Mirror symmetry in Three-Dimensional Gauge Theories, SL(2,Z) and 
D-brane Moduli Spaces}'',
Nucl.\ Phys.\ B {\bf 493} (1997) 148
[hep-th/9612131].


\bibitem{acg}
L.~Alvarez-Gaume, S.~Coleman and P.~Ginsparg,
``{\em Finiteness Of Ricci Flat N=2 Supersymmetric Sigma Models}'', 
Commun.\ Math.\ Phys.\ {\bf 103} (1986) 423.

\bibitem{hv2}
K.~Hori, S.~Katz, A.~Klemm, R.~Pandharipande, R.~Thomas,
C.~Vafa, R.~Vakil and E.~Zaslow,
CMI lectures on Mirror Symmetry.

\bibitem{DHOO}
J.~de Boer, K.~Hori, H.~Ooguri and Y.~Oz,
``{\em Mirror Symmetry in 
Three-Dimensional Gauge Theories, Quivers and D-branes}'',
Nucl.Phys. {\bf B493} (1997) 101, 
[hep-th/9611063]. 


\bibitem{sausage}
V.~A.~Fateev, E.~Onofri and A.~B.~Zamolodchikov,
``{\em The Sausage model (integrable deformations of O(3) sigma model)}'',
Nucl.\ Phys.\ B {\bf 406} (1993) 521.

\bibitem{witten} E. Witten,
``{\em Phases of N=2 Theories in Two Dimensions}'', 
Nucl.Phys. {\bf B403} (1993) 159, [hep-th/9301042].

\bibitem{morples}
D.~R.~Morrison and M.~Ronen Plesser,
``{\em Summing the Instantons:
Quantum Cohomology and Mirror Symmetry in Toric Varieties}'',
Nucl.\ Phys.\ B {\bf 440} (1995) 279
[hep-th/9412236].


\bibitem{SandS}
N.~Seiberg and S.~Shenker,
``{\em Hypermultiplet Moduli Space and String Compactification
to  Three Dimensions}'',Phys.\ Lett.\ B {\bf 388} (1996) 521, [hep-th/9608086].

\bibitem{Nikita}
N.~Nekrasov,
``{\em Five Dimensional Gauge Theories and Relativistic Integrable Systems}'',
Nucl.\ Phys.\ B {\bf 531} (1998) 323, [hep-th/9609219].

\bibitem{grass} E.~Witten,
``{\em The Verlinde Algebra and the Cohomology 
of the Grassmannian}'',
in ``Cambridge 1993, Geometry, topology, and physics'', 
357, hep-th/9312104



\bibitem{EHX}
T.~Eguchi, K.~Hori and C.~Xiong,
``{\em Gravitational Quantum Cohomology}'',
Int.\ J.\ Mod.\ Phys.\ A {\bf 12} (1997) 1743, 
[hep-th/9605225].

\bibitem{brian} B.~Greene,
``{\em String Theory on Calabi-Yau Manifolds}'', 
hep-th/9702155. 

\bibitem{mor}
D.~Morrison,
``{\em Geometric Aspects of Mirror Symmetry}'', math.AG/0007090.

\bibitem{kimyeong} H-C. Kao and K. Lee, ``{\em Self-Dual Chern-Simons Higgs 
Systems with an N=3 Extended Supersymmetry}'',  
Phys.Rev. {\bf D46} (1992) 4691, [hep-th/9205115].

\bibitem{coleman} S. Coleman and B. Hill, ``{\em No More Corrections 
to the Topological Mass Term in QED in Three Dimensions}'', 
Phys.Lett. {\bf B159} (1985) 184. 

\bibitem{seiberg} N. Seiberg, ``{\em  IR Dynamics on Branes and Space-Time 
Geometry}'', Phys.Lett. {\bf B384} (1996) 81, [hep-th/9606017].

\bibitem{HKLR}
N.~J.~Hitchin, A.~Karlhede, U.~Lindstrom and M.~Rocek,
``{\em Hyperkahler Metrics And Supersymmetry}'',
Commun.\ Math.\ Phys.\  {\bf 108} (1987) 535.

\bibitem{DHO}
J.~de Boer et al,
``{\em Dynamics of N = 2 Supersymmetric Gauge
Theories in Three Dimensions}'',
Nucl.\ Phys.\ B {\bf 500} (1997) 163
[hep-th/9703100].









\end{thebibliography}
\end{document}